\IfFileExists{/dev/null}{\immediate\write18{sh makeLinks.sh}}{\immediate\write18{makeLinks.bat}}

\documentclass[a4paper,journal,twocolumn,10pt,final]{IEEEtran}

\makeatletter
\def\subsubsection{%
	\@startsection
	{subsubsection}                 
	{3}                             
	{\z@}                           
	{2.5ex plus 1.5ex minus 1.5ex}  
	{1ex plus .5ex minus 0ex}     
	{\normalfont\normalsize\itshape}
}
\makeatother

\usepackage[none]{hyphenat}
\usepackage[utf8]{inputenc}
\usepackage{textcomp}
\usepackage{scalefnt}
\usepackage{amsmath,amsfonts,amssymb,amsbsy,amstext}
\usepackage[normalem]{ulem}
\useunder{\uline}{\ul}{}

\usepackage{mathtools}          
\usepackage{cmap}               
\usepackage[dvipsnames,svgnames,x11names]{xcolor}   
\usepackage{varwidth}

\let\originalleft\left
\let\originalright\right
\renewcommand{\left}{\mathopen{}\mathclose\bgroup\originalleft}
\renewcommand{\right}{\aftergroup\egroup\originalright}

\usepackage{graphicx}
\graphicspath{{.}{.\figs}}
\usepackage{placeins} 

\usepackage{multirow}
\usepackage{tabularx}
\newcolumntype{C}{>{\centering\arraybackslash}X}
\newcolumntype{R}{>{\flushright\arraybackslash}X}
\newcolumntype{L}{>{\flushleft\arraybackslash}X}
\newcolumntype{P}{>{\centering\arraybackslash} p{0.5\linewidth}}
\usepackage{booktabs}

\usepackage[%
style=ieee,
backend=biber, 
]{biblatex}

\defbibheading{bibliography}[\refname]{\section*{#1}}

\makeatletter
\g@addto@macro{\UrlBreaks}{\UrlOrds}
\makeatother

\usepackage[acronym,      
shortcuts,    
symbols,      
nomain,       
sanitizesort, 
nogroupskip,  
nopostdot,    
nonumberlist  
]{glossaries}[=v4.46]
\newcommand{\acro}[2]{\newacronym{#1}{#1}{#2}}

\acro{3D}{three-dimensional}
\acro{3GPP}{3rd generation partnership project}
\acro{4G}{fourth generation}
\acro{5G}{fifth generation}
\acro{ACK}{acknowledge}
\acro{AF}{amplify and forward}
\acro{AR}{autoregressive}
\acro{B5G}{beyond 5G}
\acro{BAP}{backhaul adaptation protocol}
\acro{BD}{beyond diagonal}
\acro{BLER}{block error rate}
\acro{BS}{base station}
\acro{CBR}{constant bit rate}
\acro{CDF}{cumulative distribution function}
\acro{C-link}{control link}
\acro{CN}{core network}
\acro{CNN}{convolutional neural network} 
\acro{CP}{control plane}
\acro{CQI}{channel quality indicator}
\acro{CU}{centralized unit}
\acro{DF}{decode and forward}
\acro{DL}{downlink}
\acro{DP}{data plane}
\acro{DU}{distributed unit}
\acro{EM}{electromagnetic}
\acro{ETSI}{European Telecommunications Standards Institute}
\acro{FD}{full duplex}
\acro{FL}{federated learning}
\acro{Fwd}{forwarding}
\acro{GA}{genetic algorithm}
\newacronym[longplural={next generation nodes B}]{gNB}{gNB}{next generation node B}
\acro{GR}{group report}
\acro{HAP}{high altitude platform}
\acro{HetNet}{heterogeneous network}
\acro{HD}{half duplex}
\acro{IAB}{integrated access and backhaul}
\acro{ID}{identification}
\acro{LAP}{low altitude platform}
\acro{LOS}{line of sight}
\acro{LSTM}{long short-term memory}
\acro{LTE}{long term evolution}
\acro{MAC}{medium access control}
\acro{MCS}{modulation and coding scheme}
\acro{MIMO}{multiple input multiple output}
\acro{MILP}{mixed integer linear programming}
\acro{MINLP}{mixed integer nonlinear programming}
\acro{MISO}{multiple input single output}
\acro{mmWave}{millimeter wave}
\acro{MT}{mobile termination}
\acro{NACK}{no acknowledge}
\acro{NCR}{network-controlled repeater}
\acro{NLOS}{non-line of sight}
\acro{NN}{neural network}
\acro{NR}{new radio}
\acro{OFDM}{orthogonal frequency division multiplexing}
\acro{PDCP}{packet data convergence protocol}
\acro{PHY}{physical}
\acro{PIN}{positive-intrinsic-negative}
\acro{PPP}{Poisson point process}
\acro{PRB}{physical resource block}
\acro{PSO}{particle swarm optimization}
\acro{PtP}{point-to-point}
\acro{PtmP}{point-to-multi-point}
\acro{QoS}{quality of service}
\acro{RF}{radio frequency}
\acro{RIS}{reconfigurable intelligent surface}
\acro{RL}{reinforcement learning}
\acro{RLC}{radio link control}
\acro{RR}{round robin}
\acro{RRC}{radio resource control}
\acro{SDAP}{service data adaptation protocol}
\acro{SE}{spectral efficiency}
\acro{SI}{study item}
\acro{SINR}{signal to interference-plus-noise ratio}
\acro{SISO}{single input single output}
\acro{SNR}{signal to noise ratio}
\acro{STAR}{simultaneously transmitting and reflecting}
\acro{TDD}{time division duplex}
\acro{TR}{technical report}
\acro{TS}{technical specification}
\acro{TTI}{transmission time interval}
\acro{UAV}{unmanned aerial vehicle}
\acro{UE}{user equipment}
\acro{UL}{uplink}
\acro{UP}{user plane}

\usepackage{siunitx}
\usepackage{datetime}
\usepackage{url}

\usepackage{caption}
\usepackage{subcaption}
\usepackage[linesnumbered,ruled,vlined]{algorithm2e}

\usepackage{epstopdf}

\usepackage[hidelinks=true]{hyperref}

\usepackage{enumerate}

\usepackage[final]{changes} 
\definechangesauthor[name={Victor F. Monteiro}]{vfm}
\definechangesauthor[name={Tarcisio F. Maciel}]{tfm}
\setaddedmarkup{{\color{blue}#1}}
\setdeletedmarkup{{\color{red}\sout{#1}}}

\usepackage[referable,flushleft]{threeparttablex}
\AtBeginEnvironment{tablenotes}{\footnotesize} 

\DeclareMathAlphabet{\mathppl}{T1}{ppl}{m}{it}
\DeclareMathAlphabet{\mathphv}{T1}{phv}{m}{it}
\DeclareMathAlphabet{\mathpzc}{T1}{pzc}{m}{it}

\newcommand{\Set}[1]{\mathcal{\uppercase{#1}}}

\newcommand{\stB}{\Set{B}}

\newcommand{\stR}{\Set{R}}
\newcommand{\stS}{\Set{S}}
\newcommand{\stT}{\Set{T}}
\newcommand{\stU}{\Set{U}}

\newcommand{\SecRef}[2][]{Section#1~\ref{#2}}

\newcommand{\FigRef}[2][]{Fig.#1~\ref{#2}}

\newcommand{\TabRef}[2][]{Table#1~\ref{#2}}
\newcommand{\EqRef}[2][]{\eqref{#2}}

\usepackage{standalone}
\usepackage{shapepar}

\usepackage{tikz}
\usetikzlibrary{arrows}
\usetikzlibrary{positioning}
\usetikzlibrary{shapes.misc}
\usetikzlibrary{shapes.geometric}
\usetikzlibrary{shapes.symbols}
\usetikzlibrary{external}

\usetikzlibrary{shadows}
\usetikzlibrary{backgrounds}
\usetikzlibrary{shapes}
\usetikzlibrary{shapes.multipart}
\usetikzlibrary{matrix}
\usetikzlibrary{intersections}
\usetikzlibrary{fit}
\usetikzlibrary{calc}
\usetikzlibrary{chains}
\usetikzlibrary{scopes}
\usetikzlibrary{decorations.pathreplacing}
\usetikzlibrary{decorations.text}
\usetikzlibrary{arrows.meta} 
\usetikzlibrary{mindmap}
\usetikzlibrary{trees}
\usetikzlibrary{patterns}

\usepackage{pgfplots}

\usepackage{pgfplotstable}
\usepgfplotslibrary{statistics}

\def\plotwidth{\columnwidth}
\def\plotheight{0.5\columnwidth}

\def\landscapeplotwidth{\columnwidth}
\def\landscapeplotheight{0.8\columnwidth}

\pgfplotsset{compat=1.17}

\pgfplotsset{common line style/.style={line width=1pt}}

\pgfplotsset{every axis plot post/.append style={
    every mark/.append style={scale=1.5}
}}

\pgfplotsset{common plots axis options/.style={
	every axis/.append style={
	  legend style={fill=gray!5, fill opacity=0.85, text opacity=1}
	},
	width=\plotwidth,
	height=\plotheight,
	grid=both,
	filter discard warning=false,
	tick label style={font=\footnotesize},
	label style={font=\footnotesize},
	every axis label={font=\footnotesize},
	grid=major,
	grid style={
		dashed,
		gray!30, 
		line width=0.1pt,
	},
	cycle list shift=0,
	enlargelimits={true,abs value=1pt},
	ylabel shift = -0.5ex,
	legend style={%
		font=\scriptsize,
		legend cell align=left,
		nodes={inner xsep=2pt,inner ysep=1pt,text depth=0.15em},
		/tikz/every even column/.append style={column sep = 1ex},
	},
	}
}

\pgfplotsset{bar axis options/.style={
		common plots axis options,
		ybar=1pt,
		bar width = 3pt,
		enlarge x limits={true,abs value=5pt},
}}

\pgfplotsset{mcs axis options/.style={
		bar axis options,
		ymin = 0, ymax = 100,
		xtick={0,1,...,15},
		xticklabels={{Sum of\\NACKs}, MCS 1, MCS 2, MCS 3, MCS 4, MCS 5, MCS 6, MCS 7, MCS 8, MCS 9, MCS 10, MCS 11, MCS 12, MCS 13, MCS 14, MCS 15},
		x tick label style={
			font=\tiny,
			xshift = 1ex,
			rotate=45,
			anchor=east,
			align=right,
		},
}}

\pgfplotsset{common plots axis landscape options/.style={
	every axis/.append style={
	  legend style={fill=gray!5, fill opacity=0.85, text opacity=1}
	},
	width=\landscapeplotwidth,
	height=0.8*\landscapeplotheight,
	grid=both,
	filter discard warning=false,
	tick label style={font=\footnotesize},
	label style={font=\footnotesize},
	every axis label={font=\footnotesize},
	grid=major,
	grid style={
		dashed,
		gray!30, 
		line width=0.1pt,
	},
	cycle list shift=0,
	enlargelimits={true,abs value=1pt},
	ylabel shift = -0.5ex,
	legend style={%
		font=\scriptsize,
		legend cell align=left,
		nodes={inner xsep=2pt,inner ysep=1pt,text depth=0.15em},
		/tikz/every even column/.append style={column sep = 1ex},
	},
	}
}

\pgfplotsset{bar axis landscape options/.style={
		common plots axis landscape options,
		ybar=1pt,
		bar width = 3pt,
		enlarge x limits={true,abs value=5pt},
}}

\pgfplotsset{mcs axis landscape options/.style={
		bar axis landscape options,
		ymin = 0, ymax = 100,
		xtick={0,1,...,15},
		xticklabels={{Sum of\\NACKs}, MCS 1, MCS 2, MCS 3, MCS 4, MCS 5, MCS 6, MCS 7, MCS 8, MCS 9, MCS 10, MCS 11, MCS 12, MCS 13, MCS 14, MCS 15},
		x tick label style={
			font=\tiny,
			xshift = 1ex,
			rotate=45,
			anchor=east,
			align=right,
		},
}}

\pgfplotsset{common marker style/.style={
		mark repeat = 10,
		mark size = 1pt,
		mark options={solid},
}}

\pgfplotsset{onlymacro style/.style={
	common line style,
	common marker style,
	MediumSeaGreen,
	mark=triangle*,
	densely dotted,
}}

\pgfplotsset{iab style/.style={
    common line style,
    common marker style,
    Red,
    mark=*,
    dashed,
  }}
  
\pgfplotsset{iab nlos style/.style={
    common line style,
    common marker style,
    Red,
    mark=triangle*,
    densely dotted,
  }}

\pgfplotsset{ncr style/.style={
		common line style,	
		common marker style,
		DodgerBlue,
		mark=square*,
		dashed,
}}

\pgfplotsset{ncr nlos style/.style={
		common line style,	
		common marker style,
		DodgerBlue,
		mark=triangle*,
    		densely dotted,
}}

\pgfplotsset{ris style/.style={
		common line style,	
		common marker style,
		Orange,
		mark=pentagon*,
		dashed,
}}

\pgfplotsset{uaviab style/.style={
    common line style,
    common marker style,
    Red,
    mark=diamond*,
}}

\pgfplotsset{uavncr style/.style={
	common line style,
	common marker style,
	DodgerBlue,
	mark=asterisk,
}}

\pgfplotsset{ack bar style/.style={
		DodgerBlue, fill
}}

\pgfplotsset{nack bar style/.style={
		Red, fill
}}

\pgfplotsset{ack bar two style/.style={
		DodgerBlue, pattern=north east lines,
		pattern color=DodgerBlue
}}

\pgfplotsset{nack bar two style/.style={
		Red, pattern=north east lines,
		pattern color=Red
}}

\pgfplotsset{blue style/.style={
		DodgerBlue, area legend, fill
}}

\pgfplotsset{red style/.style={
		Red, area legend, fill
}}

\pgfplotsset{solid bar style/.style={
		Black, fill
}}

\pgfplotsset{pattern bar style/.style={
		pattern=north east lines,
		pattern color=Black
}}

\pgfplotsset{common boxplot style/.style={
		line width=0.6pt,
		solid,
		mark=*,
		mark size=1pt,
}}

\pgfplotsset{onlymacro boxplot style/.style={
		common boxplot style,
		fill=MediumSeaGreen,
}}

\pgfplotsset{iab boxplot style/.style={
		common boxplot style,
		fill=Red,
}}

\pgfplotsset{ncr boxplot style/.style={
		common boxplot style,
		fill=DodgerBlue,
}}

\pgfplotsset{ris boxplot style/.style={
		common boxplot style,
		fill=Orange,
}}

\pgfplotsset{uaviab boxplot style/.style={
		common boxplot style,
		fill=Red!50!LemonChiffon,
}}

\pgfplotsset{uavncr boxplot style/.style={
		common boxplot style,
		fill=DodgerBlue!50!LemonChiffon,
}}

\def\plotsPath{figs/tikz}
\def\plotsDataPath{\plotsPath/data}

\newtoggle{tikzexternalize}
\togglefalse{tikzexternalize}

\iftoggle{tikzexternalize}{
	\immediate\write18{mkdir -p ./tikz-external-figs/}
	\usetikzlibrary{external}
	\tikzexternalize[prefix=./tikz-external-figs/,mode=list and make]
	\tikzset{external/system call={pdflatex \tikzexternalcheckshellescape -halt-on-error -interaction=batchmode -jobname "\image" "\texsource"}}
}{
}

\AtBeginDocument{%
	\abovedisplayskip 1.5ex plus 4pt minus 2pt
	\belowdisplayskip \abovedisplayskip
	\abovedisplayshortskip 0pt plus 4pt
	\belowdisplayshortskip 1.5ex plus 4pt minus 2pt
}

 \def\plotsDataPath{figs/tikz/data}

\begin{document}
\title{Cellular Network Densification: \\ a System-level Analysis with IAB, NCR and RIS}

\author{
	Gabriel C. M. da Silva,~\IEEEmembership{Student Member,~IEEE}, Victor F. Monteiro,~\IEEEmembership{Member,~IEEE}, Diego A. Sousa, \\ Darlan C. Moreira, Tarcisio F. Maciel~\IEEEmembership{Senior Member,~IEEE}, Fco. Rafael M. Lima,~\IEEEmembership{Senior Member,~IEEE} and Behrooz Makki,~\IEEEmembership{Senior Member,~IEEE}
	\thanks{Behrooz Makki is with Ericsson Research, Sweden. The other authors are with the Wireless Telecommunications Research Group (GTEL), Federal University of Cear\'{a} (UFC), Fortaleza, Cear\'{a}, Brazil. Diego A. Sousa is also with Federal Institute of Education, Science, and Technology of Cear\'{a} (IFCE), Paracuru, Brazil. This work was supported by Ericsson Research, Sweden, and Ericsson Innovation Center, Brazil, under UFC.51 Technical Cooperation Contract Ericsson/UFC. The work of Victor F. Monteiro was supported by CNPq under Grant 308267/2022-2. The work of Tarcisio F. Maciel was supported by CNPq under Grant 312471/2021-1. The work of Francisco R. M. Lima was supported by FUNCAP (edital BPI) under Grant BP5-0197-00194.01.00/22.}%
}

\maketitle

\begin{abstract}
As the number of user equipments increases in \ac{5G} and beyond, it is desired to densify the cellular network with auxiliary nodes assisting the base stations.
Examples of these nodes are \ac{IAB} nodes, \acp{NCR} and \acp{RIS}. %
In this context, this work presents a system level overview of these three nodes. %
Moreover, this work evaluates through simulations the impact of network planning aiming at enhancing the performance of a network used to cover an outdoor sport event. %
We show that, in the considered scenario, in general, \ac{IAB} nodes provide an improved signal to interference-plus-noise ratio and throughput, compared to \acp{NCR} and \acp{RIS}. %
However, there are situations where \ac{NCR} outperforms \ac{IAB} due to higher level of interference caused by the latter. %
Finally, we show that the deployment of these nodes in \acp{UAV} also achieves performance gains due to their aerial mobility. %
However, \ac{UAV} constraints related to aerial deployment may prevent these nodes from reaching results as good as the ones achieved by their stationary deployment. 
\end{abstract}

\begin{IEEEkeywords}
	IAB, NCR, RIS, UAV, 5G, network densification.
\end{IEEEkeywords}

%
\IEEEpeerreviewmaketitle
\glsresetall

\section{Introduction}
\label{SEC:intro}

The increasing number of connected devices and data traffic is pushing the boundaries of current wireless networks. 
By late~$2029$, data traffic is projected to reach $563$~exabytes per month~\cite{EricssonDataTraffic}, up from 160 exabytes~\cite{EricssonDataTraffic} at the end of 2023, marking a 3.5-fold increase. %
This surge is driven by innovations such as virtual and augmented reality, autonomous vehicles, and holographic projections. %
To accommodate this growth, \ac{5G} networks have moved beyond the congested sub-6 GHz spectrum (FR1) to include frequencies in the spectrum of \acp{mmWave} (FR2)~\cite{3gpp.38.300}. 
While FR2 offers larger bandwidth~\cite{Flamini2022}, it faces challenges such as significant path loss and difficulty penetrating obstacles~\cite{Rangan2014, Ayoubi2022}. %
Network densification with strategically placed nodes is used as a solution to extend coverage and enhance capacity. %

The primary solution to densify the network is to deploy more \acp{gNB} with fiber-based backhaul. 
However, building from scratch an entire new wired backhaul infrastructure might not be economically viable. %
Besides, it takes time, and, in certain locations such as historical areas, trenching may be prohibited~\cite{Monteiro2022}. %
In this context, wireless backhaul and technologies that increase the access extension have been considered as viable complementary solutions for achieving network densification. %

\begin{figure}[t]
	\centering
	\includegraphics[width=0.9\columnwidth]{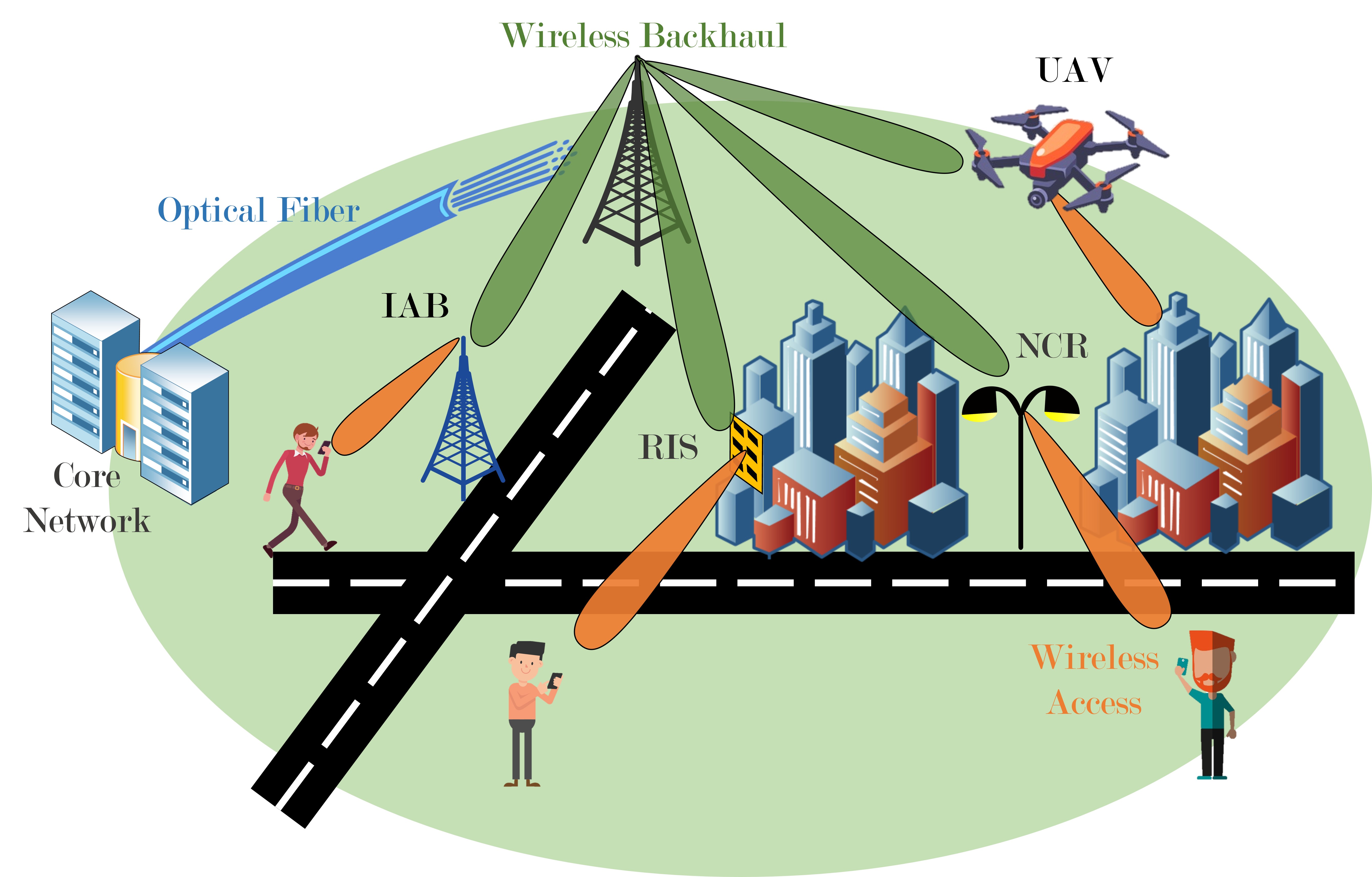}
	\caption{Illustrative scenario with \ac{IAB} node, \ac{NCR} and \ac{RIS}.}
	\label{FIG:Illustrative-scenario}
\end{figure}

Using relays for wireless backhaul is not a new concept. %
It was explored in \ac{3GPP} \ac{LTE} Release 10 with \ac{LTE} relays, but lacked commercial traction~\cite{3gpp.rel10.overview}. %
Other technologies, such as satellite and microwave \ac{PtP}/\ac{PtmP}, also use wireless backhaul~\cite{Jaber2016}. %
Despite its history, wireless backhaul now garners increased commercial interest due to its proved ability to achieve over \SI{100}{Gbps}~\cite{Czegledi2020} and its flexibility and cost-effectiveness~\cite{Tezergil2022}. %
Currently, a node with wireless backhaul under consideration is \ac{IAB} nodes, and nodes which enhance the access extension are \acp{NCR} and \acp{RIS}~\cite{Flamini2022}. %
Figure~\ref{FIG:Illustrative-scenario} illustrates a scenario featuring these nodes.


\ac{IAB} is a multi-hop setup where the same hardware and/or spectrum is used for both backhauling and access communication to the \acp{UE}. 
This means that both access and backhaul links are wireless. %
\ac{IAB} nodes began to be standardized by \ac{3GPP} in Release 16~\cite{3gpp.38.874}, which addressed possible architectures, radio protocols, and physical layer aspects. %
Notably, access and backhaul links can be either in-band, sharing the same frequency spectrum, or out-of-band, using separate frequency bands. %


Concerning \acp{NCR}, they were introduced by \ac{3GPP} in Release 18~\cite{3gpp.38.867}. %
\acp{NCR} can be seen as an enhancement of traditional \ac{RF} repeaters. 
One of the key differences between an \ac{NCR} and a traditional \ac{RF} repeater is that an \ac{NCR} has beamforming capability controlled by a serving \ac{gNB} via a side control link~\cite{3gpp.38.867}.
Furthermore, unlike a \ac{RF} repeater, \ac{NCR} is fully aware of the \ac{DL}/\ac{UL} split in the \ac{TDD} scheme, and is under the control of its controlling \ac{gNB} which provides the \ac{NCR} with side information~\cite{qualcomm2021}. %

With respect to \acp{RIS}, they are made up of multiple antenna patches that use passive elements to introduce phase shifts to incoming signals~\cite{etsi.GR.RIS.001, Astrom2024}. %
More specifically, they are composed of a metamaterial layer controlled by a \ac{gNB}. %
Specifically, the metamaterial is a surface consisting of reflecting elements with electrical thickness in the order of subwavelength of the signal to which it is designed to work with. %
This allows for programmable control of the wireless environment~\cite{Wu2021}.  %

To improve the performance of such nodes, a \ac{LOS} link is recommended. %
This challenge should be managed during deployment, but achieving \ac{LOS} between a serving \ac{gNB} and an \ac{IAB}/\ac{NCR}/\ac{RIS} can be difficult in certain situations, such as in urban areas with tall buildings. 
One temporary solution to overcome this problem is to mount these nodes on \acp{UAV}, which also offer mobility. %
It is important to note that, while \ac{3GPP} has standardized both stationary and mobile \ac{IAB} nodes, \acp{NCR} are only standardized for stationary use, and \ac{RIS} lacks standardization. %

Few studies simultaneously address these three types of nodes. %
In \cite{Flamini2022}, the authors refer to the environment where these technologies coexist as a heterogeneous smart electromagnetic environment. %
They provide an overview of each node, their industrial progress and their standardization status. %
However, they do not provide simulation results comparing the nodes' performances. %
They evaluate only the \ac{RIS} performance. %
Another work addressing these nodes is \cite{Wen2024}. %
The authors review each technology, discuss the \ac{3GPP} standardization status of \ac{IAB} and \ac{NCR}, and present reports from the \Ac{ETSI} on \ac{RIS}. %
However, as~\cite{Flamini2022}, \cite{Wen2024} lacks simulation results comparing these nodes. %
Furthermore, system-level analyses are crucial. %
The presence of multiple of such nodes might have resource allocation and interference issues which can not be captured in link-level analyses, as presented in other works.


In this paper, we provide an overview of \ac{IAB}, \ac{NCR} and \ac{RIS}. %
We discuss their main characteristics, e.g., architecture and specific features. %
Furthermore, we address different levels of mobility supported by them, including the possibility of deployment on \acp{UAV}. %
In this context, a brief overview of \acp{UAV} as serving nodes is also presented. %
A literature review of related research topics is provided for each node. %
We also analyze in a system-level perspective the impact of deploying these nodes in the environment. %
Simulation results are presented to assess the performance of these nodes when deployed either as stationary nodes or mounted on mobile \acp{UAV}. %



The present work is organized as follows. %
First, \SecRef{SEC:HetNet:Nodes} provides a conceptual overview of \ac{IAB}, \ac{NCR}, \ac{RIS} and \ac{UAV}, including their architectural details and a literature review. %
Next, \SecRef{SEC:SYS_MODEL} introduces a system level model allowing a performance evaluation of these nodes. %
Based on the presented model and \ac{3GPP} standardized procedures, \SecRef{SEC:performance_evaluation} presents simulation results to assess the impact of deploying these nodes on system performance. %
An extensive performance analysis is then presented. %
Finally, \SecRef{SEC:CONC} summarizes the conclusions of the work. %


\section{Emerging Wireless Network Nodes Used to Serve \acp{UE}}
\label{SEC:HetNet:Nodes}

\subsection{Integrated Access and Backhaul}
\label{SEC:HetNet:IAB}

The standardization of \ac{IAB} by \ac{3GPP} started as a \ac{SI} of \ac{3GPP} Release~15 in 2017. %
In Release~16, a \ac{TR}~\cite{3gpp.38.874} published in 2018 detailed \ac{IAB} system level aspects, e.g., topology management for single/multi-hop and redundant connectivity, route selection and dynamic resource allocation between the access and backhaul links. 
Later on, in 2020, Release~16 also started specifying \ac{IAB} on two \acp{TS}: \ac{TS}~38.300~\cite{3gpp.38.300}, specifying protocol layers and architecture of \ac{IAB}, and \ac{TS}~38.174~\cite{3gpp.38.174}, presenting the minimum \ac{RF} characteristics and minimum performance requirements of \ac{IAB}. %
In 2021, it was created, in Release~17, a work item entitled ``Enhancements to \ac{IAB} for NR''~\cite{3gpp.rp.210758}. %
The main enhancements were related to: 1) duplexing (\ac{IAB} was constrained to \ac{HD} mode in Release 16, but it envisioned to support \ac{FD} mode); 2) topology adaptation (procedures for inter-donor \ac{IAB} node migration to enhance robustness and load balancing); and 3) routing and transport (improve multi-hop latency and congestion mitigation). %



\begin{figure}[t]
	\centering
	\includegraphics[width=\columnwidth]{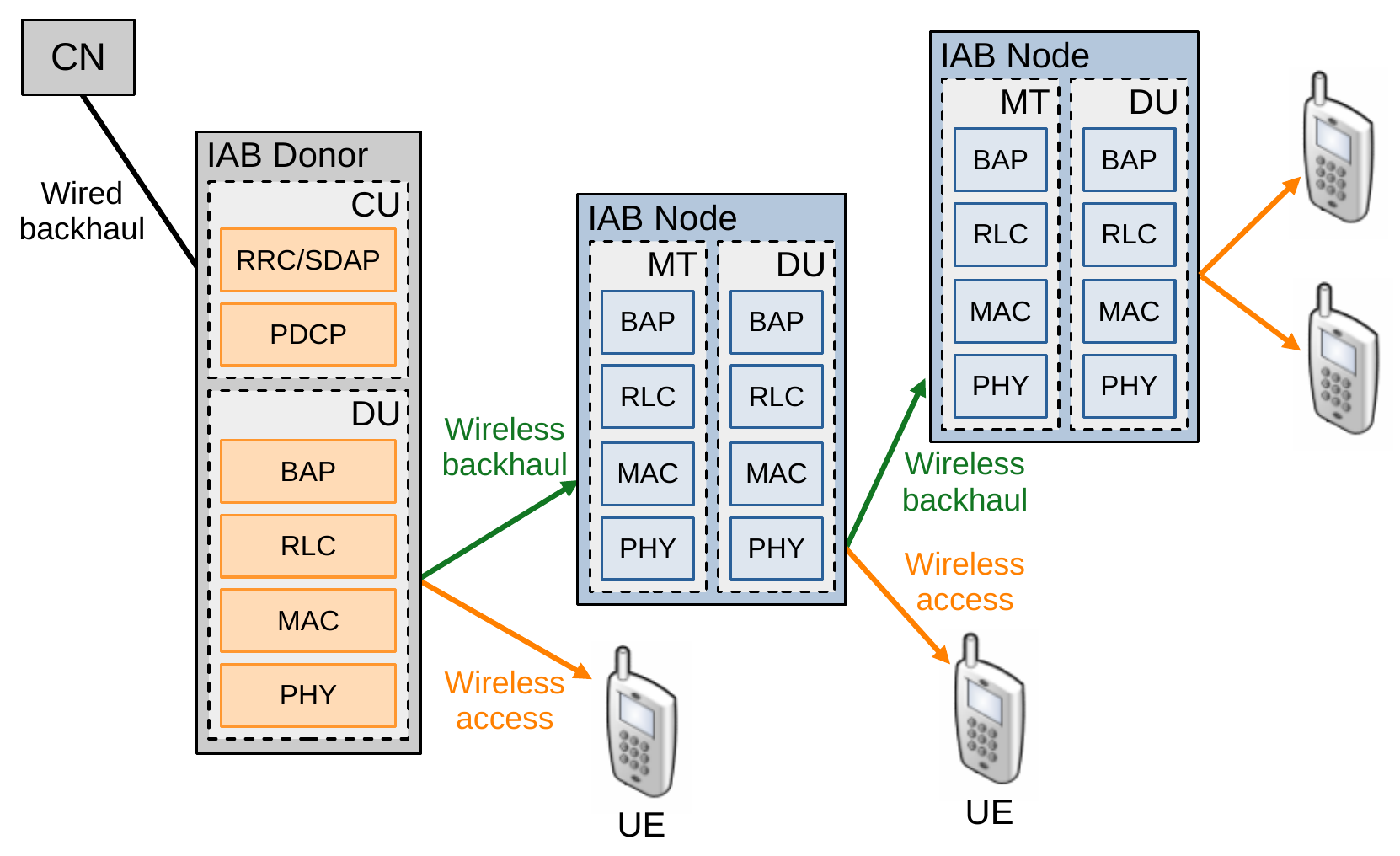}
	\caption{Network with \ac{IAB} nodes.}
	\label{FIG:IAB:iab_system}
\end{figure}

Figure~\ref{FIG:IAB:iab_system} illustrates a network where some of the serving nodes have a wireless backhaul. %
The serving nodes in a system with \ac{IAB} can be classified into two types: \ac{IAB} donor and \ac{IAB} node. %
The \ac{IAB} donor is connected to the \ac{CN} by a non-\ac{IAB}, e.g., fiber backhaul, while the \ac{IAB} nodes are connected to the \ac{IAB} donor or another \ac{IAB} node by a wireless \ac{IAB}-based backhaul. %
The \acp{UE}, depicted as mobile phones, can be served by both \ac{IAB} node types. %


\ac{IAB} donor can be split in two parts transparent to the served nodes~\cite{3gpp.38.300}: \ac{CU} and \ac{DU}. %
As in a regular \ac{gNB}, on the one hand, the \ac{CU} is responsible for less time-critical radio functionalities. %
On the other hand, the \ac{DU} part is responsible for time-critical radio functionalities, e.g., scheduling and retransmission. %
This split is related to the possibility of deploying both parts in different geographical locations. %
For example, the \ac{DU} may be close to the served nodes, in order to reduce communication delay, and the \ac{CU} may be in a place with higher processing capability. %
Regarding their protocols, \ac{DU} is responsible for lower layers, e.g., \ac{PHY}, \ac{MAC} and \ac{RLC}; while \ac{CU} is responsible for higher layers, e.g., \ac{PDCP} and \ac{RRC}/\ac{SDAP}. %

%

Regarding the \ac{IAB} node, it is split in \ac{DU} and \ac{MT}. %
From a \ac{UE} perspective, the \ac{DU} acts as a regular \ac{gNB}. %
From the perspective of an \ac{IAB} donor, the \ac{MT} acts as a regular \ac{UE}. %


To handle the traffic between an \ac{IAB} donor and multiple \ac{IAB} nodes, it was introduced a new sublayer called \ac{BAP}~\cite{3gpp.38.340}. %
The \ac{IAB} donor configures each \ac{IAB} node with an unique \ac{BAP} \ac{ID} and a routing table that indicates its parent and possible children. 

Taking into account the presented technical features of \ac{IAB}, many works have already investigated different aspects related to this topic. %
In the following, we present some of them. %

An important aspect usually addressed is the impact of \ac{IAB} node position in the system performance. %
Some of the works that have evaluated this point are~\cite{Madapatha2020, Madapatha2021, Carvalho2023}. %

In~\cite{Madapatha2020}, the authors consider a scenario where \acp{UE} are distributed in the coverage area of small and macro \acp{BS}. 
The backhaul link of the small \acp{BS} can be via fiber or wireless (in that case, it is an \ac{IAB} node). %
Furthermore, the \ac{IAB} network is submitted to blockage, tree foliage and rain. %
The authors conclude that it is necessary to have more \ac{IAB} nodes than fiber-connected \acp{gNB} to achieve similar performance (in terms of \acp{UE} minimum required rate). %


In~\cite{Madapatha2021} and~\cite{Carvalho2023}, \acp{GA} are utilized to place \ac{IAB} nodes. %
The authors from~\cite{Madapatha2021} consider a scenario with temporary blockages and optimize via \ac{GA} the \ac{IAB} position to increase the service coverage probability. %
In~\cite{Carvalho2023}, the authors study the cost-benefit between a scenario with only macro \acp{BS} and another with \ac{IAB} nodes. %

Another topic related to \ac{IAB} systems is topology adaptation/routing, addressed in~\cite{Polese2020, Tran2023, Ghodhbane2023}. %
The authors of~\cite{Polese2020} compare two types of path selection policies, i.e., highest-quality-first and wired-first. %
The highest-quality-first policy chooses as parent node the candidate \ac{BS} with the highest \ac{SNR}, while the wired-first policy prefers choosing a fiber-connected \ac{BS}, i.e., an \ac{IAB} donor, instead of another \ac{IAB} node, even if the \ac{IAB} donor provides a channel link worse than the one provided by another \ac{IAB} node. %
The paper shows that, in the considered scenario, the wired-first policy outperforms the other strategy in terms of latency due a reduced number of intermediary nodes. 

In~\cite{Tran2023} and~\cite{Ghodhbane2023}, the authors propose algorithms for topology adaptation and routing aiming at load balancing. %
In~\cite{Tran2023}, the proposed solution is based on packet queue's length and uses graph theory to optimize the \ac{IAB} network. %
The proposed algorithm increases the average throughput and the percentage of satisfied \acp{UE}.  %
In \cite{Ghodhbane2023}, the algorithm aims at optimizing the trade-off between spectral efficiency and load balancing. %
In heavy data loading, the proposed solution outperforms benchmark solutions in terms of system rate. %
For light data loading, all solutions have similar results. %


A third topic usually addressed by \ac{IAB} works is the in-band resource allocation of access and backhaul links. %
Examples of works on this topic include \cite{Lei2020, Lai2020, Sana2023}. %

In~\cite{Lei2020}, the authors formulate a problem of spectrum resource allocation aiming at maximizing the sum log-rate of all \ac{UE}. %
Due to the challenge of finding a solution for a \ac{MINLP} problem, they propose a deep reinforcement learning method to reach real-time spectrum allocation satisfying the system requirements. %

In~\cite{Lai2020}, the authors' goal is to maximize the rate of the \acp{UE}. %
For this, they propose optimal and suboptimal algorithms to multiplex the \acp{UE} in frequency and space. %
They consider a power constraint scenario. %

The objective of~\cite{Sana2023} is to simultaneously solve three problems where the scenario has mobile \ac{IAB} nodes. %
The problems are: 1) define \ac{IAB} nodes positions; 2) define \ac{UE}-\ac{IAB} association; 3) in-band resource allocation of backhaul and access links.  %
To solve them, the authors propose a two-level hierarchical multi-agent learning, i.e., a type of \ac{RL}. 

\subsection{Network Controlled Repeater}
\label{SEC:HetNet:NCR}

In previous generations of wireless cellular networks, e.g., the \ac{4G}, \ac{RF} repeaters were used to enhance network coverage. %
They work as \ac{AF} relays, introducing negligible latency, since they neither decode nor process the forwarded signal. %
Moreover, they are cost-effective and easy to deploy. %
However, they forward unwanted interference signals due to their lack of signal processing capability~\cite{qualcomm2021, 3gpp.38.867}, and they do not perform adaptive beamforming towards \acp{UE} which is required specially in FR2~\cite{3gpp.RP-210818}. %

\Ac{NCR} is an evolution of the traditional \ac{RF} repeater. %
It was standardized in \ac{3GPP} Release~18. 
It can be controlled by the network via side control link, which enables, for example, an \ac{NCR} to operate with spatial diversity via beamforming controlled by the network~\cite{3gpp.38.867, Xin2022}. %

Besides the fact that \acp{NCR} operate as \ac{AF} repeaters, while \ac{IAB} nodes operate as \ac{DF} relays, there are other differences between them. %
For example, on the one hand, \ac{NCR} was standardized to be deployed as a stationary single-hop node. %
On the other hand, \ac{IAB} can operate as a mobile node in a multi-hop deployment. %
The \ac{NCR} is fully controlled by its controlling \ac{gNB}, while the \ac{IAB} node can make decisions on its own. 
Furthermore, \ac{IAB} is more expensive than \ac{NCR}.

\begin{figure}[t]
	\centering
	\includegraphics[width=\columnwidth]{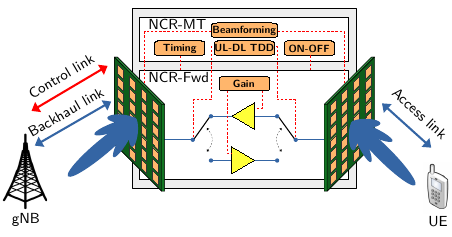}
	\caption{\ac{NCR} architecture.}
	\label{FIG:NCR:ncr_architecture}
\end{figure}

An \ac{NCR} can be split into two functional entities, i.e., \ac{NCR}-\ac{MT} and \ac{NCR}-\ac{Fwd}, as shown in \FigRef{FIG:NCR:ncr_architecture}. %
The \ac{NCR}-\ac{MT} entity is used to communicate with its controller \ac{gNB} through a \ac{C-link}. 
The side control information controls the actions of \ac{NCR}-\ac{Fwd}. %
Regarding the \ac{NCR}-\ac{Fwd}, it is responsible for amplifying and forwarding the \ac{RF} signals between a \ac{gNB} and its serving \acp{UE}~\cite{Silva2024a}. %

A common \ac{NCR} deployment consists of having at least two antenna arrays, where one is responsible for receiving/transmisting control and backhaul links, and another is responsible for the access links. %

Some of the information that can be exchanged through the \ac{C-link} are~\cite{Wen2024}:

\begin{itemize}
	\item \textit{Beamforming information:} Traditional \ac{RF} repeaters transmit omnidirectionally, causing interference and providing less power to target \acp{UE}. In contrast, \ac{NCR} utilizes adaptive beamforming to address severe losses in \ac{mmWave}, with the beamforming configuration managed by the controller \ac{gNB} through \ac{C-link}.
	
	
	\item \textit{Timing information:} \ac{gNB} can configure the temporal alignment between \ac{NCR} reception and transmission for \ac{DL} and \ac{UL}. %
	Specifically, a \ac{gNB} may configure an \ac{NCR} to introduce a delay in communication to prevent issues like interference between sequential transmissions. 
	
	\item \textit{Information on \ac{UL}-\ac{DL} \ac{TDD} configuration:} \ac{C-link} is also used to configure the \ac{NCR} with a \ac{TDD} scheme. %
	\ac{NCR} supports semi-static \ac{TDD} \ac{DL}/\ac{UL} for access, backhaul, and control links to reduce the signaling. %
	The \ac{DL}/\ac{UL} \ac{TDD} configuration is the same for access and backhaul links~\cite{Xin2022}.%
	
	\item \textit{ON-OFF information:}	This information is sent to \ac{NCR} for turning on/off the \ac{NCR}-\ac{Fwd} part. %
	The main goals of this functionality is to save power and to mitigate potential interference caused by an \ac{NCR}.  
		The default mode is OFF. %
	It stays in this state until a different \ac{gNB} indication~\cite{3gpp.38.867}. %
\end{itemize}

State-of-the-art works cover different aspects of \ac{NCR}. %
In~\cite{Xin2022}, the authors provide an overview on \ac{NCR}. %
They present general aspects of side control information and discuss mechanisms for \ac{NCR} identification and authentication. %
They compare three deployments: 1) macro only; 2) macro + \ac{RF} repeater; and 3) macro + \ac{NCR}. %
They also evaluate the impact of beamforming information on \ac{NCR}'s performance. %

The authors of \cite{Carvalho2024} discuss \acp{NCR}' main specifications in \ac{3GPP} Release~18 and the challenges associated with \ac{NCR}-assisted networks, such as the lack of testbeds and cost-efficiency trade-offs between \ac{IAB} and \ac{NCR}. %
Their simulation results suggest that deploying \ac{NCR} can enhance performance particularly for \ac{UL} and cell-edge \acp{UE}. %


In \cite{Leone2022, Ayoubi2023, Guo2022}, the authors compare \ac{NCR} and \ac{RIS}. %
Work~\cite{Leone2022} formulates mixed integer linear problems to maximize network reliability based on \ac{NCR} and \ac{RIS} positions and configurations. %
Study~\cite{Ayoubi2023} focuses on physical layer and compares \ac{NCR} and \ac{RIS} coverage. %
It concludes that \ac{RIS} is preferable for open areas despite its lower capacity, while \ac{NCR} is better for narrow scenarios, e.g., roads. %
The authors of \cite{Guo2022} evaluate \ac{NCR} with and without power constraints and \ac{RIS} with two beamforming algorithms, including hardware impairments. %
They conclude that, in general, \ac{NCR} outperforms \ac{RIS}, with \ac{RIS} suffering the most when hardware impairments are present. %


The authors of \cite{Dong2023} evaluate an \ac{NCR} with two panels for the access link. %
They aim to optimize a multi-user precoder to maximize the number of served \acp{UE}, considering constraints on the number of time-frequency resources and on the minimum required rate per \ac{UE}. %
Compared with \acp{NCR} with only one antenna array in the access, two panels doubled the \ac{SE} and the number of served \acp{UE}. %


In~\cite{Sousa2024}, the authors study the problem of beam squinting in \ac{NCR} from link and system level perspectives. %
They consider signaling (i.e., \ac{CP}) and data (i.e., \ac{UP}) transmissions in different subbands. 
They propose a compensation method to prevent beam squinting caused by the configuration of the \ac{NCR} to transmit data (\ac{UP}) in a given subband based on measurements (\ac{CP}) performed in another subband. %
The results show that, with their proposal, the \ac{NCR} performs as good as if \ac{CP} and \ac{UP} shared the same subbands. %


In~\cite{Silva2023a}, the authors analyze the interference impact of the \ac{NCR} in a neighbor cell. %
They conclude that the interference can be mitigated by the choice of the beam that is used to serve the \acp{UE}. %
Besides, placing an \ac{NCR} closer to the serving \ac{UE} is not necessarily the position that maximizes the served \ac{UE} performance nor the one that most harms a neighbor cell. %


\subsection{Reconfigurable Intelligent Surface}
\label{SEC:HetNet:RIS}

Regarding \ac{3GPP} work on \ac{RIS}, up to this moment, it has not started an \ac{SI} on this topic. %
There was an expectation that an \ac{SI} would be discussed in Releases~$18$ and~$19$, however it was not confirmed. %
Concerning the \Ac{ETSI}, in~$2023$, a \ac{GR} was formulated to deliberates about several aspects related to \acp{RIS}, e.g., deployment scenarios, architectures, and communication model~\cite{etsi.GR.RIS.001, etsi.GR.RIS.002, etsi.GR.RIS.003}. %

\ac{RIS} can dynamically or semi-statically alter \ac{EM} properties of an incident wave. %
It consists of a thin metamaterial layer typically controlled by a \ac{gNB}. 
This allows control over reflection, refraction, absorption and backscattering effects~\cite{etsi.GR.RIS.001}, enabling the creation of \ac{LOS} links. %
The metamaterial surface has tunable properties, electronically adjusted by devices like \acp{PIN} or varactor diodes~\cite{Wu2021}.  %
A \ac{RIS} standalone deployment is also envisioned with an integrated controller. %
A \ac{RIS} transforms the wireless environment from a passive entity into a programmable intelligent agent~\cite{etsi.GR.RIS.001}. %


A \ac{gNB} manages the \ac{RIS} by collecting data from the \ac{RIS} and the \acp{UE} to estimate their channels and compute optimal beamforming. %
The \ac{gNB} then configures the \ac{RIS} to adjust its properties to improve system performance. %
Notably, the \ac{gNB} handles all data processing, while the \ac{RIS} merely implements the instructions received, which reduces production costs by eliminating the need for an internal processing unit. %

\begin{figure}[t]
	\centering
	\includegraphics[width=\columnwidth]{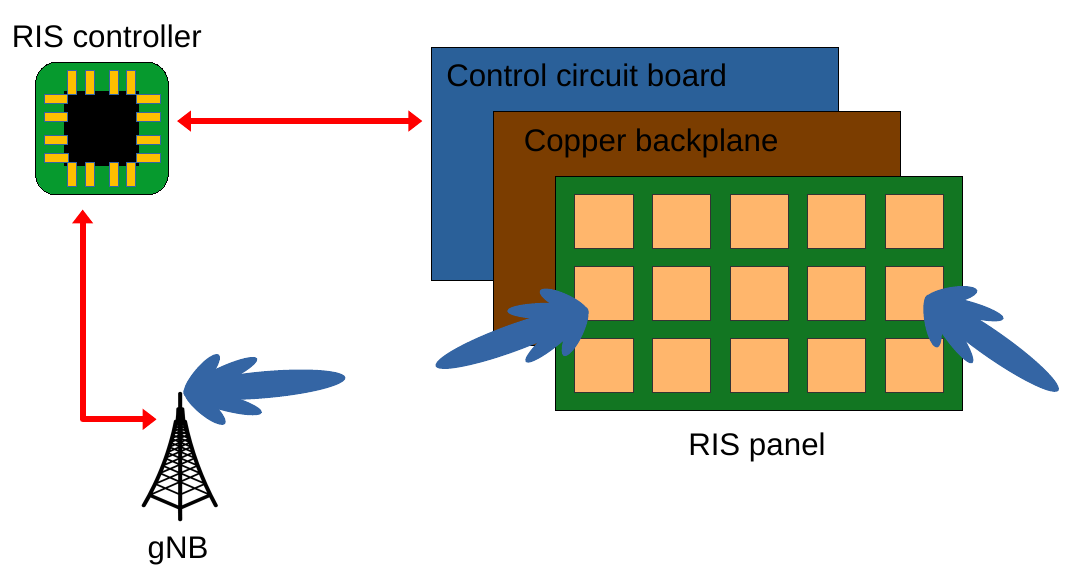}
	\caption{RIS architecture.}
	\label{FIG:RIS:RIS_architecture}
\end{figure}

As shown in~\FigRef{FIG:RIS:RIS_architecture}, a \ac{RIS} is composed of two main components: the \ac{RIS} controller and the \ac{RIS} panel. %
The \ac{RIS} controller adjusts the properties of the unit cells to manipulate incoming \ac{EM} waves based on \ac{gNB} control signals. 
The panel is structured in multiple layers, with a simple model featuring three layers. %
The inner layer contains a control circuit connected to the \ac{RIS} controller, which powers the circuit and manages its response. %
The middle layer, typically made of copper, serves as a shield to prevent signal leakage. %
The outermost layer consists of metallic unit cells that interact with the incident \ac{EM} waves modifying its properties, e.g., amplitude, phase, and polarization~\cite{etsi.GR.RIS.002}. %



\ac{RIS} research spans various areas, with early studies focusing on channel estimation, as explored in~\cite{Gomes2023,Guo2023,Molazadeh2022,Ginige2024}. %
In \cite{Gomes2023}, researchers explore a communication system aided by \ac{RIS}, considering physical imperfections, e.g., hardware-induced phase shift variations. %
They propose two tensor-based channel estimation algorithms using parallel factor analysis and evaluate their computational complexity. %

The authors of \cite{Guo2023} identify two key limitations in existing channel estimation algorithms: 1) rapid increase in computational complexity with more \ac{RIS} elements, and 2) reliance on specific matrix structures. %
They propose an algorithm based on unitary approximate message passing, which scales linearly with the number of \ac{RIS} elements and does not depend on specific matrix structures. %



The authors of \cite{Ginige2024} consider a \ac{RIS}-assisted system and propose a channel estimation method using convolutional neural network and autoregressive predictors. 
Compared to their benchmark, they achieve high prediction accuracy, improved spectral efficiency, and reduced pilot overhead. %


\ac{RIS} channel modeling is addressed in \cite{Tang2021, Meng2024, Chen2024, Liu2024}. %
In \cite{Tang2021}, it is formulated a closed-form expression for free-space path loss models. %
Some aspects that are analyzed are the far-field and near-field beamforming. %
To validate their theoretical findings, they present measurements performed during an experiment in an anechoic chamber with three \acp{RIS}. %
In \cite{Meng2024} and \cite{Chen2024}, \acp{RIS} are deployed in mobile \acp{UAV}. %
Both works study the channel statistics in the considered scenario. %
In \cite{Liu2024}, it is introduced a method for calculating the scattered electric field generated by a \ac{RIS} considering the mutual coupling of the unit-cells. %

Beam design and selection is another well-researched area related to \ac{RIS}-assisted networks, with notable works including \cite{Ibrahim2023, Munawar2023, Amiriara2023}. %
Paper~\cite{Ibrahim2023} surveys beamforming design, identifies research gaps, and suggests future research directions. %
Work~\cite{Munawar2023} proposes a low-complexity adaptive beam selection scheme. %
It involves using two sub-optimal beamforming vectors, calculating the \ac{RIS} beamforming, and selecting the solution that maximizes received signal strength. %
In~\cite{Amiriara2023}, the authors consider the problem of maximizing the sum-rate of all \acp{UE}. %
They propose two solutions: a three-step convex optimization and a neural network scheme. %

It is important to highlight that \ac{RIS} is a huge area and the present work does not have the intention to cover all the topics. %
Other areas that can be mentioned are simultaneously transmitting and reflecting \ac{RIS}~\cite{Mu2022}, liquid-crystal \ac{RIS}~\cite{Ndjiongue2021} and beyond diagonal \ac{RIS}~\cite{Li2023}. %
\subsection{Unmanned Aerial Vehicle}
\label{SEC:HetNet:UAV}

In recent years, \ac{UAV} applications have expanded into various fields, including delivery services, rescue, surveillance, military operations, and telecommunications~\cite{Mozaffari2019}. 
This increased interest is due to \acp{UAV}' advantages unmatched by traditional ground-based networks, e.g., fast \ac{3D} deployment and maneuverability. 

\acp{UAV} offer several benefits, as described in \cite{Li2019b,Dai2023}. %
Their altitudes increase the likelihood of establishing \ac{LOS} links with ground \acp{gNB}, resulting in more reliable data transmission compared to terrestrial networks with potential obstructions. %
Additionally, \acp{UAV} can adjust their positions to improve signal quality and provide dynamic deployment capabilities. %
They can reposition autonomously to meet real-time network needs, enhancing network robustness and adaptability to changing conditions. %
\acp{UAV} can be rapidly deployed in emergencies, e.g., natural disasters, without the need for site rentals or extensive setup. %
They can form swarms to quickly expand network coverage. %
However, due to battery limitations, \acp{UAV} are best suited for temporary solutions, making them ideal for events and disaster response. %

\acp{UAV} can operate mainly as aerial \acp{BS}/relays or as aerial \acp{UE}. %
In the aerial base station/relay role, \acp{UAV} provide reliable and fast-deployable wireless communication. %
Conversely, \acp{UAV} functioning as aerial \acp{UE} are suited for data collection, delivery tasks, and surveillance~\cite{Banafaa2024}. %
%

\acp{UAV} can also be categorized by their operational altitude~\cite{Mozaffari2019}. %
In this class, we can mention at least two categories: \ac{HAP} and \ac{LAP}. %
\acp{HAP} operate above $20$~km and remain quasi-stationary for extended periods, while \acp{LAP} fly at lower altitudes, typically up to tens of meters, and are valued for their rapid deployment and mobility~\cite{Karabulut2021}. %



Recognizing the potential of \acp{UAV}, the \ac{3GPP} has undertaken extensive studies on \ac{UAV} integration. %
\ac{TR}~22.829~\cite{3gpp.22.829} explores \ac{UAV} support for various low altitudes applications, while \ac{TR}~36.777~\cite{3gpp.36.777} focuses on their integration with \ac{LTE} networks. %
Additionally, \ac{TR}~23.754~\cite{3gpp.23.754} and \ac{TS}~23.256~\cite{3gpp.23.256} examine mechanisms for \ac{UAV} identification, tracking, authorization, and authentication. %

In the following, we present some works that focus on the synergy between \acp{UAV} and networks with  wireless backhaul. %
Notable works in the area of \ac{UAV}-\ac{IAB} integration include~\cite{Tafintsev2020a, Zhang2022}. %
In \cite{Tafintsev2020a}, the authors examine the system-level impact of deploying \acp{UAV} with \ac{IAB} on mobile \ac{UE} performance. %
They compare static and mobile \ac{UAV}-\ac{IAB} setups and optimize \ac{UAV} mobility using a \ac{PSO} algorithm. %
Three backhaul scenarios are considered: 1) backhaul-unaware (limited bandwidth not considered in optimization); 2) backhaul-aware (limited bandwidth considered in optimization); and ideal (unlimited bandwidth). %
The study assesses \ac{UE} throughput and Jain's fairness index for various numbers of deployed \acp{UAV}. %


In \cite{Zhang2022}, the authors explore a public safety scenario, where a \ac{UAV}-\ac{IAB} node covers an area affected by infrastructure failure. %
The \ac{UAV}-\ac{IAB} node, which operates in the mid-band frequency, receives signals from a nearby \ac{BS}, offering broader coverage compared to \ac{mmWave} bands. %
The study proposes a \ac{RL} solution to optimize the \ac{UAV}-\ac{IAB} node's \ac{3D} positioning and antenna tilt. %
They also outline the required network architecture and signaling procedures for deploying their solution. %
System-level simulations show that their approach improves throughput and reduces \ac{UE} drop rates. %



To the best of our knowledge, no research has yet documented the mounting of an \ac{NCR} onto a \ac{UAV}. %
This gap indicates potential for exploring \ac{UAV}-\ac{NCR} technology. %


The works \cite{Park2022, Yang2023} investigate using \ac{RIS} mounted on \acp{UAV} to enhance wireless communication. %
In~\cite{Park2022}, a method is proposed for optimizing the \ac{RIS} phase vector and bandwidth allocation for each \ac{UE} using \ac{LSTM} and federated learning. %
This two-stage approach involves the \ac{UAV} first training a model to predict the phase vector and then using the \ac{RIS} to maximize the total achievable data rate. %

%

In \cite{Yang2023}, the authors propose a model combining Q-learning and Fourier feature mapping for a \ac{RIS}-mounted \ac{UAV} system. %
This model optimizes \ac{UAV}-\ac{RIS} positioning, transmission beamforming, and \ac{RIS} reflection beamforming. %
The goal is to enhance communication \ac{QoS}, reliability, and security against eavesdroppers and jammers. 




\section{System Model}
\label{SEC:SYS_MODEL}
%

\begin{figure}[t]
	\centering
	\includegraphics[width=0.9\columnwidth]{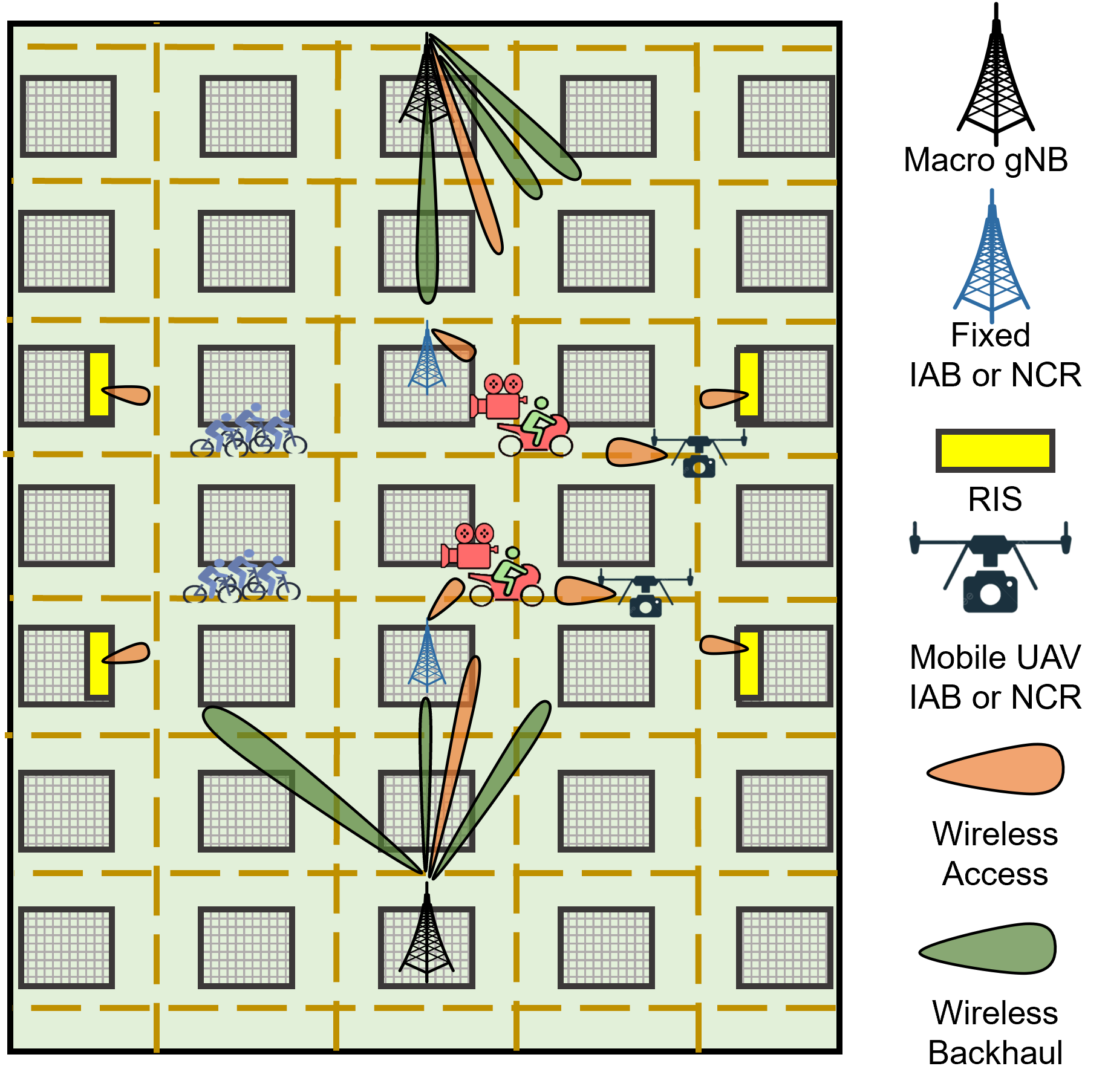}
	\caption{Illustration of the possible deployments to serve the cameramen on motorcycles: only \acp{gNB}, stationary \ac{IAB}/\ac{NCR}/\ac{RIS}, and \ac{UAV} mounted \ac{IAB}/\ac{NCR} nodes.}
	\label{FIG:SYS_MODEL:scenario}
\end{figure}

Based on use case 5.11.2.2 ``User outside the vehicle'' of~\cite{3gpp.22.839}, a scenario is considered where an outdoor event takes place, i.e., a bike race. %
As shown in~\FigRef{FIG:SYS_MODEL:scenario}, a simplified Madrid grid scenario~\cite{METIS:D6.1:2013} is considered. %
Cameramen on motorcycles broadcast live videos of the race. %
Half of them are positioned on one street and the other, on a parallel street. %


In this scenario, six different deployments are considered. %
The first deployment includes only two traditional macro \acp{gNB} (black nodes in \FigRef{FIG:SYS_MODEL:scenario}). %
The second, third, and fourth deployments respectively add stationary \ac{IAB} nodes, stationary \acp{NCR}, or stationary \acp{RIS} to the traditional macro \acp{gNB} (with \ac{IAB} and \ac{NCR} represented by blue nodes and \ac{RIS} by yellow nodes in \FigRef{FIG:SYS_MODEL:scenario}). %
The \ac{RIS} is strategically placed at a different position to take advantage of its reflective properties. %
The fifth and sixth scenarios consider the deployment of \ac{IAB} nodes or \acp{NCR} mounted on mobile \acp{UAV} to follow the cameramen, in addition to the traditional macro \acp{gNB}. %


In the second and third deployments, the stationary \ac{IAB} node and \ac{NCR} are positioned two blocks away from the \ac{gNB}, as shown in \FigRef{FIG:SYS_MODEL:scenario}. %
Both nodes are equipped with three antenna arrays each for improved street coverage. %
These arrays are spaced $120^{\circ}$ apart to reduce interference. %
One array, referred to as the backhaul antenna array, faces the \ac{gNB}, while the other two, known as access antenna arrays, are directed towards the street where the cameramen are located, with $120^{\circ}$ between them. %
In contrast, the mobile \ac{UAV} is equipped with only two antenna arrays to minimize weight: one for the backhaul link and one for the access links. %


In the fourth deployment, the \acp{RIS} are positioned differently from the stationary \ac{IAB} nodes and \acp{NCR} due to their reflection properties. %
Four \acp{RIS} are used instead of two to enhance coverage and match the number of access antenna arrays of the stationary \ac{IAB} nodes and \acp{NCR}. %


The following subsections present the \ac{SINR} expressions for \acp{UE} connected via an \ac{IAB} node, \ac{NCR}, or \ac{RIS}. %
These equations are formulated for the \ac{DL}, though a similar approach applies to the \ac{UL}. %
Additionally, \ac{IAB} operates in in-band mode. %
The available frequency band is divided into $K$ \acp{PRB}, where each \ac{PRB} consists of a number of adjacent subcarriers. %


\subsection{\ac{SINR} of \acp{UE} Served by \ac{IAB} Nodes}

The useful power $S_{u_{x}, k}^{IAB}$ received by \ac{UE} $u_{x}$ served by an \ac{IAB} node, e.g., \ac{IAB} node $r_x$, at \ac{PRB} $k$ can be expressed as:

\begin{equation}
	S_{u_{x}, k}^{IAB} = \gamma_{r_x, u_{x}, k} \cdot p_{r_x, u_{x}, k},
\end{equation}

\noindent where $p_{r_{x}, u_{x}, k}$ is transmit power used by \ac{IAB} node $r_x$ to transmit useful signal to \ac{UE} $u_{x}$ at \ac{PRB} $k$; and $\gamma_{i,j,k} = \textbf{d}_{j,k} \textbf{H}_{i,j,k} \textbf{f}_{i,k}$ denotes the combined effect of the channel $\textbf{H}_{i,j,k}$ after the transmission and reception filters $\textbf{f}_{i,k}$ and $\textbf{d}_{j,k}$, respectively, applied to a signal transmitted by node $i$ to node $j$ at \ac{PRB} $k$.

The interference signals on \ac{UE} $u_{x}$ comes from two types of sources: backhaul links and access links. %
More precisely, signals belonging to the first type are transmitted at \ac{PRB} $k$ by a given \ac{IAB} node, e.g., \ac{IAB} node $r_i$, to its serving \ac{gNB} $b_i$. %
Moreover, signals belonging to the second type are transmitted in the access links for other \acp{UE}, e.g., \ac{UE} $u_{y}$, at \ac{PRB} $k$. %
The transmitter $q_y$ is its serving node, which can be either another \ac{IAB} or a macro \ac{gNB}. %
Thus, the interference $I_{u_{x}, k}^{IAB}$ perceived by \ac{UE} $u_{x}$ at \ac{PRB} $k$ is equal to:

\begin{equation}
\begin{aligned}
I_{u_{x}, k}^{IAB} =& \sum_{\substack{(r_{i}, b_{i}) \in \stR \times \stB}} \gamma_{r_i, u_x, k} \cdot p_{r_i, b_i, k} \\
&+ \sum_{\substack{(q_y, u_y) \in  \left(\stR \cup \stB\right) \times \stU, \\ y \neq x}} \gamma_{q_{y}, u_{x}, k} \cdot p_{q_y, u_{y}, k} ,
\end{aligned}
\end{equation}

\noindent where $\stR$, $\stB$ and $\stU$ are the sets of \acp{IAB} nodes, \acp{gNB} and \acp{UE} in the system, respectively. %

Finally, the \ac{SINR} $\rho_{u_{x}, k}^{IAB}$ perceived by \ac{UE} $u_{x}$ at \ac{PRB} $k$ is 

\begin{equation}
	\rho_{u_{x}, k}^{IAB} = \frac{S_{u_{x}, k}^{IAB}}{I_{u_{x}, k}^{IAB} + \sigma^{2}_{k}},
	\label{EQ:SYS_MODEL:SINR_IAB_DL}
\end{equation}

\noindent where $\sigma^{2}_{k}$ is the noise component at \ac{PRB} $k$.

\subsection{\ac{SINR} of \acp{UE} Served through \acp{NCR}}

It is neither considered the link between \acp{NCR} nor the \ac{NCR} self-interference. %
The former can be mitigated during the deployment by avoiding overlapping coverage areas. %
The latter can be mitigated by placing the \ac{NCR} antenna arrays at an angular distance of $120$~degrees or more~\cite{Dong2023}. %



Consider a \ac{UE} $u_{x}$ connected to \ac{gNB} $b_x$ through \ac{NCR} $s_x$. %
Its useful power received  at \ac{PRB} $k$ can be split into two parts, i.e., the \ac{gNB} direct link and the link through the \ac{NCR}:

\begin{equation}
\begin{aligned}
	S_{u_{x}, k}^{NCR} = \gamma_{b_{x}, u_{x}, k} &\cdot p_{b_{x}, u_{x}, k} \\
	&+ \gamma_{s_{x}, u_{x}, k} \cdot g_{s_{x}, k} \cdot \gamma_{b_{x}, s_{x}, k} \cdot p_{b_{x}, u_{x}, k},
\end{aligned}
\label{EQ:SYS_MODEL:S_NCR_DL}
\end{equation}

\noindent where $p_{b_{x}, u_{x}, k}$ is the power transmitted by \ac{gNB} $b_{x}$, and $g_{s_{x}, k}$ is the stationary power gain applied by \ac{NCR} $s_{x}$ at \ac{PRB} $k$. %


The interference can be split into three sets: 1) signals received directly from other \acp{gNB}, e.g., \ac{gNB} $b_i$ with $i \neq x$, and transmitted at \ac{PRB} $k$ to  others \acp{UE} $u_y$; 2) copies of the signals that belong to the first set and that were amplified by an \ac{NCR}; and 
3) amplified noises from all \acp{NCR} with exception of the one \ac{UE} $u_{x}$ is connected through. %
Thus, the interference $I_{u_{x}, k}^{NCR}$ perceived by \ac{UE} $u_{x}$ at \ac{PRB} $k$ is equal to:


\begin{equation}
	\begin{aligned}
		I_{u_{x}, k}^{NCR} =& \sum_{\substack{(u_y, b_y) \in \stU \times \stB, \\ y \neq x}} \gamma_{b_y, u_x, k} \cdot p_{b_y, u_y, k} \\
		&+ \sum_{\substack{(u_y, b_y) \in \stU \times \stB, \\ y \neq x}} \sum_{\substack{s_i \in \stS}} \gamma_{s_i, u_x, k} \cdot g_{s_i, k} \cdot \gamma_{b_y, s_i, k} \cdot p_{b_y, u_y, k} \\ 
		&+ \sigma^{2}_{k} \cdot \sum_{\substack{s_{i} \in \stS, \\ y \neq x}} \gamma_{s_{i}, u_{x}, k} \cdot g_{s_{i}, k}, \\
		\label{EQ:SYS_MODEL:I_NCR_DL}
	\end{aligned}
\end{equation}

\noindent where $\stS$ is the set of \acp{NCR}.

The overall noise power $N_{u_{x}, k}$ received by \ac{UE} $u_{x}$ at \ac{PRB} $k$ is the noise component plus the noise amplified by the \ac{NCR} to which \ac{UE} $u_{x}$ is connected to. It can be expressed as:


\begin{equation}
	N_{u_{x}, k} = \sigma^{2}_{k} ( 1 + \gamma_{s_{x}, u_{x}, k} \cdot g_{s_{x}, k} ).
	\label{EQ:SYS_MODEL:N_NCR_DL}
\end{equation}

Finally, the \ac{SINR} $\rho_{u_{x}, k}^{NCR}$ perceived by \ac{UE} $u_{x}$ at \ac{PRB} $k$ is

\begin{equation}
	\rho_{u_{x}, k}^{NCR} = \frac{S_{u_{x}, k}^{NCR}}{I_{u_{x}, k}^{NCR} + N_{u_{x}, k}}.
	\label{EQ:SYS_MODEL:SINR_NCR_DL}
\end{equation}

\subsection{\ac{SINR} of \acp{UE} Served through \acp{RIS}}

Consider a \ac{UE} $u_{x}$ connected to \ac{gNB} $b_x$ through \ac{RIS} $t_x$. %
The useful power $S_{u_{x}, k}^{RIS}$ received  at \ac{PRB} $k$ by \ac{UE} $u_{x}$ can be split into two parts as in the \ac{NCR} case, i.e., the direct link from the \ac{gNB} and the link through the \ac{RIS}: 

\begin{equation}
	S_{u_{x}, k}^{RIS} = \gamma_{b_{x}, u_{x}, k} \cdot p_{b_{x}, u_{x}, k} + \eta_{b_{x}, u_{x}, k} \cdot p_{b_{x}, u_{x}, k},
\end{equation}

\noindent where $\eta_{i,j,k} = \textbf{d}_{j,k} \textbf{H}_{t_{x},j,k} \boldsymbol{\Theta} \textbf{H}_{i,t_{x},k} \textbf{f}_{i,k}$ denotes the combined effect of the channel when signal is received through \ac{RIS}. %
$\textbf{H}_{i,t_{x},k}$ is the channel between transmitter $i$, whose precoder is $\textbf{f}_{i,k}$, and the \ac{RIS} $t_{x}$ at \ac{PRB} $k$. %
The matrix of reflection coefficients is given by $\boldsymbol{\Theta}$ and its calculation is based on \cite{Munawar2023}, which is an algorithm optimizes the reflection coefficients and \ac{gNB} beamforming for maximizing the signal strength received by the \ac{UE}. %
$\textbf{H}_{t_{x}, j,k}$ is channel between \ac{RIS} $t_{x}$ and receiver $j$, whose combiner is $\textbf{d}_{j,k}$ at \ac{PRB} $k$. %

As in the \ac{NCR} case, we neither consider the \ac{RIS} self-interference nor interference between \acp{RIS}. %
In practice, \ac{RIS} depends on \ac{LOS} link in outdoor urban settings, where the likelihood of \ac{RIS}-\ac{RIS} links is minimal due to obstacles. %

Similarly to~\EqRef{EQ:SYS_MODEL:I_NCR_DL}, except for the third term representing amplified noise, the interference received by \ac{UE} $u_x$ is: %

\begin{equation}
	\begin{aligned}
		I_{u_{x}, k}^{RIS} =& \sum_{\substack{(u_y, b_y) \in \stU \times \stB, \\ y \neq x}} \gamma_{b_y, u_x, k} \cdot p_{b_y, u_y, k} \\
		&+ \sum_{\substack{(u_y, b_y) \in \stU \times \stB, \\ y \neq x}} \sum_{\substack{t_i \in \stT}} \eta_{b_{y}, u_{x}, k} \cdot p_{b_y, u_y, k}, \\
	\end{aligned}
\end{equation}

\noindent where $\stT$ is the set of \acp{RIS}.

Finally, the \ac{SINR} $\rho_{u_{x}, k}^{RIS}$ perceived by \ac{UE} $u_{x}$ at \ac{PRB} $k$ is 

\begin{equation}
	\rho_{u_{x}, k}^{RIS} = \frac{S_{u_{x}, k}^{RIS}}{I_{u_{x}, k}^{RIS} + \sigma^{2}_{k}},
	\label{EQ:SYS_MODEL:SINR_RIS_DL}
\end{equation}

\noindent where the $\sigma^{2}_{k}$ is the noise component at \ac{PRB} $k$.

\section{Performance Evaluation}
\label{SEC:performance_evaluation}
%

This section compares the performance of a scenario with only macro cells (referred to as the macro only scenario) against scenarios that include either stationary or mobile nodes (mounted on a UAV) with \ac{IAB}, \ac{NCR}, or \ac{RIS}. %
Details on the used parameters are provided in \SecRef{SEC:performance_evaluation:Sim_Assumptions}, and the results are discussed in \SecRef{SEC:performance_evaluation:Sim_Res}. %


\subsection{Simulation Assumptions}
\label{SEC:performance_evaluation:Sim_Assumptions}

\begin{table*}[!t]
	\centering
	\caption{Entities characteristics.}
	\label{TABLE:performance_evaluation:Entities-characteristics}
	\begin{tabularx}{\textwidth}{lXXXXX}
		\hline
		\textbf{Parameter} & \textbf{Macro \ac{gNB}} & \textbf{Stationary \ac{IAB}/\ac{NCR}} & \textbf{Stationary \ac{RIS}} & \textbf{UAV \ac{IAB}/\ac{NCR}} & \textbf{\ac{UE}} \\
		\hline
		Height & 25 m & 10 m & 40 m & 40 m & 1.5 m \\
		Transmit power & 35 dBm & 32 dBm & - & 29 dBm & 24 dBm \\
		Antenna array & URA $8\times 8$ & URA $4\times 4$ (3 panels) & URA $8\times 8$ & URA $4\times 4$ (2 panels) & Single Antenna \\
		Antenna elem. pattern & \ac{3GPP} 3D~\cite{3gpp.38.901} & \ac{3GPP} 3D~\cite{3gpp.38.901} & \ac{3GPP} 3D~\cite{3gpp.38.901} & \ac{3GPP} 3D~\cite{3gpp.38.901} & Omni \\
		Max. antenna elem. gain & 8 dBi & 8 dBi & 8 dBi & 8 dBi & 0 dBi \\
		Speed & \SI{0}{km/h} & \SI{0}{km/h} & \SI{0}{km/h} & \SI{40}{km/h} & \SI{40}{km/h} \\
		\hline
	\end{tabularx}
\end{table*}

\begin{table}[!t]
	\caption{Characteristics of the links in same cell.}
	\begin{center}
		{
			\begin{tabular}{ccc}\hline
				\textbf{Link in same cell}  				& \textbf{Scenario}  								& \textbf{LOS/NLOS} \\\hline
				gNB - UE       								& Urban Macro~\cite{3gpp.38.901c}   				& NLOS \\
				gNB - Stationary \ac{IAB}/\ac{NCR}/\ac{RIS}      & Urban Macro~\cite{3gpp.38.901c}\cite{3gpp.36.777} & LOS\\
				gNB - Mobile \ac{UAV} w/ \ac{IAB}/\ac{NCR}  & Urban Macro~\cite{3gpp.36.777}    				& LOS \\
				\ac{IAB}/\ac{NCR}/\ac{RIS} - UE  			& Urban Micro~\cite{3gpp.38.901c}\cite{3gpp.36.777}	& LOS \\%
				Mobile \ac{UAV} w/ \ac{IAB}/\ac{NCR} - UE   & Urban Micro~\cite{3gpp.36.777}    				& LOS \\%
				UE - UE  									& Urban Micro~\cite{3gpp.38.901c}					& LOS \\\hline
			\end{tabular}
			\begin{tabular}{ccc}\hline
				\textbf{Link in different cell}				& \textbf{Scenario}  								& \textbf{LOS/NLOS} \\\hline
				gNB - UE       								& Urban Macro~\cite{3gpp.38.901c}   				& NLOS \\
				gNB - Stationary \ac{IAB}/\ac{NCR}/\ac{RIS}      & Urban Macro~\cite{3gpp.38.901c}\cite{3gpp.36.777} & NLOS \\
				gNB - Mobile \ac{UAV} w/ \ac{IAB}/\ac{NCR}  & Urban Macro~\cite{3gpp.36.777}    				& LOS \\
				\ac{IAB}/\ac{NCR}/\ac{RIS} - UE  			& Urban Micro~\cite{3gpp.38.901c}\cite{3gpp.36.777}	& NLOS \\%
				Mobile \ac{UAV} w/ \ac{IAB}/\ac{NCR} - UE   & Urban Micro~\cite{3gpp.36.777}    				& LOS \\%
				UE - UE  									& Urban Micro~\cite{3gpp.38.901c}					& NLOS \\\hline
			\end{tabular}
		}
	\end{center}
	\label{TABLE:performance_evaluation:LINKS}
\end{table}

\begin{table}[!t]
	\centering
	\setlength{\tabcolsep}{1ex}
	\caption{Simulation parameters.}
	\label{TABLE:performance_evaluation:Simul_Param}
	\begin{tabularx}{0.99\columnwidth}{>{\raggedright\arraybackslash}X>{\raggedright\arraybackslash}X}
		\hline
		\textbf{Parameter} & \textbf{Value} \\
		\hline
		Carrier frequency & 28 GHz\\
		System bandwidth & \SI{50}{MHz}\\
		Subcarrier spacing & 60 kHz\\
		Number of subcarriers per \acs{PRB} &  $12$\\
		Number of \acsp{PRB} & $66$\\
		Slot duration & 0.25 ms \\
		OFDM symbols per slot & $14$ \\
		Channel generation procedure & As described in~\cite[Fig.7.6.4-1]{3gpp.38.901}\\
		Path loss  & Eqs. in~\cite[Table 7.4.1-1]{3gpp.38.901}\\
		Fast fading & As described in~\cite[Sec.7.5]{3gpp.38.901} and \cite[Table 7.5-6]{3gpp.38.901} \\
		AWGN density power per subcarrier & -174 dBm/Hz\\
		Noise figure &  9 dB\\
		Number of \acp{UE} & 8 \\
		Traffic model & Full buffer \\
		NCR gain & 60 dB \\
		\hline
	\end{tabularx}
\end{table}

As presented in~\SecRef{SEC:SYS_MODEL}, it was considered a simplified Madrid grid (\FigRef{FIG:SYS_MODEL:scenario}) consisting of 120~m $\times$ 120~m blocks with 3~m wide sidewalks and 14~m wide streets. %
The \acp{UE}, i.e., cameramen's equipments, were initially placed in the left block, with half in the upper middle street and the other half in the lower middle street, as depicted in \FigRef{FIG:SYS_MODEL:scenario}. %
Additionally, the \acp{UE} moved through 3 blocks at a speed of 40~km/h. %


The adopted channel model was based on the \ac{3GPP} standards outlined in~\cite{3gpp.38.901c} and~\cite{3gpp.36.777}. %
It featured a carrier frequency of $28$~GHz with bandwidth of $50$~MHz. %
The model accounted for spatial and temporal consistency, distance-dependent path loss, lognormal shadowing, and small-scale fading. %
Details of our channel model implementation are provided in \cite{Pessoa2019}. %
Other channel characteristics are listed in Tables~\ref{TABLE:performance_evaluation:LINKS} and~\ref{TABLE:performance_evaluation:Simul_Param}. %


In the time domain, the minimum scheduling unit was a slot of $0.25$~ms with 14~\ac{OFDM} symbols. %
The considered \ac{TDD} scheme alternated \ac{DL} and \ac{UL} transmissions between slots. %
Regarding the frequency domain, one \ac{PRB} consisted of $12$~consecutive subcarriers with subcarrier distance of $60$~kHz. %
The \ac{PRB} allocation was based on the \ac{RR} scheduler. %


Concering the link adaptation, a target \ac{BLER} of 10\% was set using the \ac{CQI}/\ac{MCS} mapping curves standardized in~\cite{3gpp.38.214}. %
An outer loop strategy was considered to avoid the increase of the \ac{BLER}. %
It decreased the estimated \ac{SINR} by $1$~dB after transmission errors and increased by $0.1$~dB after error-free transmissions.
Tables~\ref{TABLE:performance_evaluation:Entities-characteristics} and \ref{TABLE:performance_evaluation:Simul_Param} present other simulation parameters. %

\subsection{Simulation Results}
\label{SEC:performance_evaluation:Sim_Res}

Figures~\ref{FIG:performance_evaluation:SINR_down} and \ref{FIG:performance_evaluation:SINR_up} present the \ac{CDF} of the \ac{SINR} in \ac{DL} and \ac{UL}, respectively, as detailed in Eqs.~\eqref{EQ:SYS_MODEL:SINR_IAB_DL}, \eqref{EQ:SYS_MODEL:SINR_NCR_DL} and \eqref{EQ:SYS_MODEL:SINR_RIS_DL} for \ac{IAB}, \ac{NCR} and \ac{RIS}, respectively. %
In the baseline scenario, i.e., the macro only scenario (dotted green curves), the \acp{UE} experienced lower \ac{SINR}, particularly in the \ac{UL}, with $80$\% of \acp{UE} connections having \ac{SINR} below \SI{2.6}{dB}. %
This was due to low \ac{UE} transmit power and the distance between \ac{gNB} and \acp{UE}, i.e., exceeding \SI{300}{m}. %
This highlights the need of improving \ac{UL} links. %
Furthermore, we can see that deploying \ac{IAB} nodes and \acp{NCR} as either stationary or mounted on mobile \acp{UAV} and \ac{RIS} improved the \ac{SINR} of the \acp{UE} in both \ac{UL} and \ac{DL} compared to macro only. %

\begin{figure}[t]
	\centering
		\begin{tikzpicture}
		\begin{axis}[common plots axis options,
			ylabel = CDF (\%),
			xlabel = SINR (dB),
			ymin = 0, ymax = 100,
			xmin = -10, xmax = 60,
			xtick = {-10,0,...,60},
			legend style={
		    	at = {(0.5, 1.05)},
		    	anchor = south,
		    	legend columns = 3,
			}
			]
			
			\def\isDownlink{True}
			
			\pgfplotstableread [col sep=comma] {\plotsDataPath/SINR/sinr_cdf_NCR_0_backhaul_los_True_isDownlink_\isDownlink.csv}\tableDataOnlyMacro
			
			\pgfplotstableread [col sep=comma] {\plotsDataPath/SINR/sinr_cdf_NCR_1_2_backhaul_los_True_isDownlink_\isDownlink.csv}\tableDataNCR
			\pgfplotstableread [col sep=comma] {\plotsDataPath/SINR/sinr_cdf_IAB_1_2_backhaul_los_True_isDownlink_\isDownlink.csv}\tableDataIAB
			
			\pgfplotstableread [col sep=comma] {\plotsDataPath/SINR/sinr_cdf_UAVIAB_1_2_backhaul_los_True_isDownlink_\isDownlink.csv}\tableDataUAVIAB
			\pgfplotstableread [col sep=comma] {\plotsDataPath/SINR/sinr_cdf_UAVNCR_1_2_backhaul_los_True_isDownlink_\isDownlink.csv}\tableDataUAVNCR
			
			\pgfplotstableread [col sep=comma] {\plotsDataPath/SINR/sinr_cdf_RIS_2_88_backhaul_los_True_isDownlink_\isDownlink.csv}\tableDataRIS

			\addlegendimage{iab style}
			\addlegendentry{Stationary IAB}
			
			\addlegendimage{ncr style}
			\addlegendentry{Stationary NCR}
			
			\addlegendimage{onlymacro style}
			\addlegendentry{Macro Only}
			
			\addlegendimage{uaviab style}
			\addlegendentry{UAV IAB}
			
			\addlegendimage{uavncr style}
			\addlegendentry{UAV NCR}
			
			\addlegendimage{ris style}
			\addlegendentry{Stationary RIS}

			\addplot[onlymacro style]
			table[x=x, y expr = \thisrow{y}*100] from \tableDataOnlyMacro;
			
			\addplot[iab style]
			table[x=x, y expr = \thisrow{y}*100] from \tableDataIAB;
			
			\addplot[ncr style]
			table[x=x, y expr = \thisrow{y}*100] from \tableDataNCR;
			
			\addplot[ris style]
			table[x=x, y expr = \thisrow{y}*100] from \tableDataRIS;
			
			\addplot[uaviab style]
			table[x=x, y expr = \thisrow{y}*100] from \tableDataUAVIAB;
			
			\addplot[uavncr style]
			table[x=x, y expr = \thisrow{y}*100] from \tableDataUAVNCR;
		\end{axis}
	\end{tikzpicture}
	%

	\caption{\ac{CDF} of \ac{SINR} in \ac{DL}.}
	\label{FIG:performance_evaluation:SINR_down}
\end{figure}
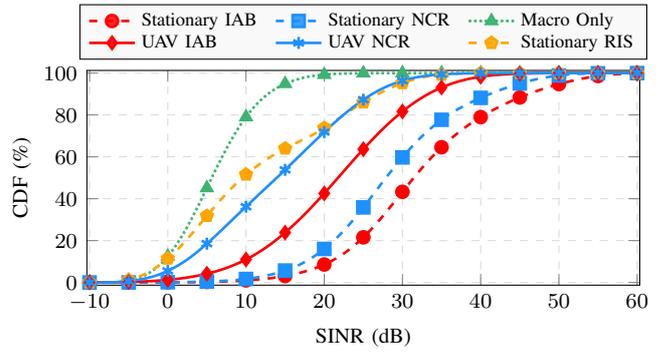

\begin{figure}[t]
	\centering
		\begin{tikzpicture}
		\begin{axis}[common plots axis options,
			ylabel = CDF (\%),
			xlabel = SINR (dB),
			ymin = 0, ymax = 100,
			xmin = -10, xmax = 60,
			xtick = {-10,0,...,60},
			legend style={
		    	at = {(0.5, 1.05)},
		    	anchor = south,
		    	legend columns = 3,
			}
			]
			
			\def\isDownlink{False}
			
			\pgfplotstableread [col sep=comma] {\plotsDataPath/SINR/sinr_cdf_NCR_0_backhaul_los_True_isDownlink_\isDownlink.csv}\tableDataOnlyMacro
			
			\pgfplotstableread [col sep=comma] {\plotsDataPath/SINR/sinr_cdf_NCR_1_2_backhaul_los_True_isDownlink_\isDownlink.csv}\tableDataNCR
			\pgfplotstableread [col sep=comma] {\plotsDataPath/SINR/sinr_cdf_IAB_1_2_backhaul_los_True_isDownlink_\isDownlink.csv}\tableDataIAB
			
			\pgfplotstableread [col sep=comma] {\plotsDataPath/SINR/sinr_cdf_UAVIAB_1_2_backhaul_los_True_isDownlink_\isDownlink.csv}\tableDataUAVIAB
			\pgfplotstableread [col sep=comma] {\plotsDataPath/SINR/sinr_cdf_UAVNCR_1_2_backhaul_los_True_isDownlink_\isDownlink.csv}\tableDataUAVNCR
			
			\pgfplotstableread [col sep=comma] {\plotsDataPath/SINR/sinr_cdf_RIS_2_88_backhaul_los_True_isDownlink_\isDownlink.csv}\tableDataRIS

			\addlegendimage{iab style}
			\addlegendentry{Stationary IAB}
			
			\addlegendimage{ncr style}
			\addlegendentry{Stationary NCR}
			
			\addlegendimage{onlymacro style}
			\addlegendentry{Macro Only}
			
			\addlegendimage{uaviab style}
			\addlegendentry{UAV IAB}
			
			\addlegendimage{uavncr style}
			\addlegendentry{UAV NCR}
			
			\addlegendimage{ris style}
			\addlegendentry{Stationary RIS}

			\addplot[onlymacro style]
			table[x=x, y expr = \thisrow{y}*100] from \tableDataOnlyMacro;
			
			\addplot[iab style]
			table[x=x, y expr = \thisrow{y}*100] from \tableDataIAB;
			
			\addplot[ncr style]
			table[x=x, y expr = \thisrow{y}*100] from \tableDataNCR;
			
			\addplot[ris style]
			table[x=x, y expr = \thisrow{y}*100] from \tableDataRIS;
			
			\addplot[uaviab style]
			table[x=x, y expr = \thisrow{y}*100] from \tableDataUAVIAB;
			
			\addplot[uavncr style]
			table[x=x, y expr = \thisrow{y}*100] from \tableDataUAVNCR;
		\end{axis}
	\end{tikzpicture}
	%

	\caption{\ac{CDF} of \ac{SINR} in \ac{UL}.}
	\label{FIG:performance_evaluation:SINR_up}
\end{figure}
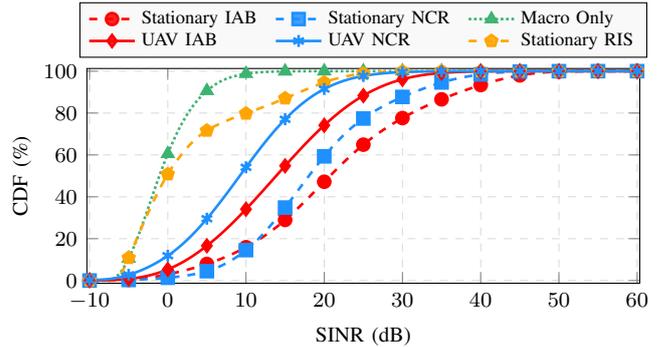

In scenarios with stationary nodes (dashed curves), \ac{IAB} performed better than \ac{NCR}, and \ac{NCR} outperformed \ac{RIS}. %
This is because the \ac{SINR} for \ac{NCR} accounts for backhaul link quality (Eq.~\eqref{EQ:SYS_MODEL:SINR_NCR_DL}), while \ac{IAB} does not (Eq.~\eqref{EQ:SYS_MODEL:SINR_IAB_DL}). %
In the results shown, \ac{NCR} benefited from a \ac{LOS} backhaul link (\TabRef{TABLE:performance_evaluation:LINKS}). %
Performance differences between stationary \ac{NCR} and \ac{IAB} are more pronounced with \ac{NLOS} backhaul, as will be discussed later. %
This underscores the importance of a \ac{LOS} backhaul for \ac{NCR} deployments. %


In the \ac{UL} (\FigRef{FIG:performance_evaluation:SINR_up}), \acp{UE} with lower \ac{SINR} experienced slightly better \ac{SINR} with \ac{NCR} compared to \ac{IAB}. %
This is because \ac{IAB} introduces interference in \ac{UE} \ac{UL} transmissions due to its backhaul simultaneously operating in \ac{DL}, unlike \ac{NCR}, where both backhaul and access links operated in \ac{UL}. 

%

Comparing \ac{NCR} and \ac{RIS}, unlike \ac{NCR}, the \ac{RIS} did not amplify the \ac{SINR}, it merely reflected the signal. %
This underscores the importance of \ac{LOS} propagation for \ac{RIS} to deliver beneficial outcomes. %

In the \ac{RIS} scenario, the lower part of the \ac{SINR} curves resembles the macro-only case, while the higher part shows improved \ac{SINR} values. %
This is because the \ac{RIS} was optimized to point towards a specific \ac{UE}, benefiting its neighbors but not others farther. %
Thus, \acp{UE} not within the \ac{RIS} optimization range communicated directly with the macro \ac{BS}, leading to similar low \ac{SINR} values as the macro-only scenario. %


Comparing the scenarios with \ac{IAB} and \ac{NCR} mounted on a \ac{UAV}, \ac{IAB} performed better than \ac{NCR} in both \ac{DL} and \ac{UL}. %
Although the \ac{NCR} was expected to benefit from \ac{LOS} backhaul, this did not compensate for its lower performance. %
Stationary nodes consistently achieved higher \ac{SINR} values compared to mobile \ac{UAV} nodes. %
In \ac{DL}, this was due to stationary nodes having twice the transmit power of \ac{UAV} nodes. %
In \ac{UL}, stationary nodes benefited from shorter distances to \acp{UE}. %
Additionally, stationary nodes had double the number of access panels compared to mobile \ac{UAV} nodes, which only had one access panel. %
These constraints in \acp{UAV}' transmit power and number of access panels are related to the \acp{UAV}' capability and size limitations compared to stationary nodes. %

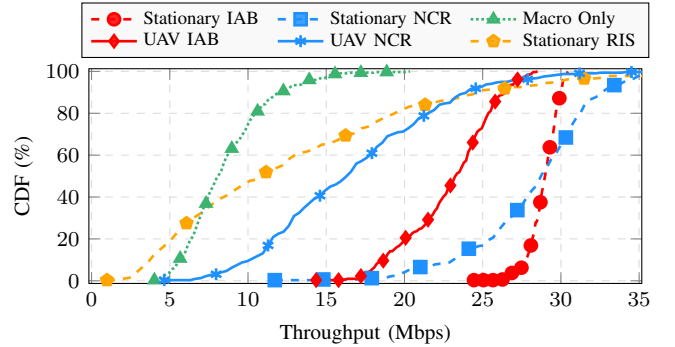
\begin{figure}[t]
	\centering
		\begin{tikzpicture}
		\begin{axis}[common plots axis options,
			ylabel = CDF (\%),
			xlabel = Throughput (Mbps),
			ymin = 0, ymax = 100,
			xmin = -0.001, xmax = 35,
			xtick = {0,5,...,35},
			legend style={
		    	at = {(0.5, 1.05)},
		    	anchor = south,
		    	legend columns = 3,
			}
			]
			
			\def\isDownlink{True}
			
			\pgfplotstableread [col sep=comma] {\plotsDataPath/Throughput/e2e_throughput_mbps_cdf_NCR_0_backhaul_los_True_isDownlink_\isDownlink.csv}\tableDataOnlyMacro
			\pgfplotstableread [col sep=comma] {\plotsDataPath/Throughput/e2e_throughput_mbps_cdf_NCR_1_2_backhaul_los_True_isDownlink_\isDownlink.csv}\tableDataNCR
			
			\pgfplotstableread [col sep=comma] {\plotsDataPath/Throughput/e2e_throughput_mbps_cdf_IAB_1_2_backhaul_los_True_isDownlink_\isDownlink.csv}\tableDataIAB
			\pgfplotstableread [col sep=comma] {\plotsDataPath/Throughput/e2e_throughput_mbps_cdf_RIS_2_88_backhaul_los_True_isDownlink_\isDownlink.csv}\tableDataRIS
			
			\pgfplotstableread [col sep=comma] {\plotsDataPath/Throughput/e2e_throughput_mbps_cdf_UAVIAB_1_2_backhaul_los_True_isDownlink_\isDownlink.csv}\tableDataUAVIAB
			\pgfplotstableread [col sep=comma] {\plotsDataPath/Throughput/e2e_throughput_mbps_cdf_UAVNCR_1_2_backhaul_los_True_isDownlink_\isDownlink.csv}\tableDataUAVNCR

			\addlegendimage{iab style}
			\addlegendentry{Stationary IAB}
			
			\addlegendimage{ncr style}
			\addlegendentry{Stationary NCR}
			
			\addlegendimage{onlymacro style}
			\addlegendentry{Macro Only}
			
			\addlegendimage{uaviab style}
			\addlegendentry{UAV IAB}
			
			\addlegendimage{uavncr style}
			\addlegendentry{UAV NCR}
			
			\addlegendimage{ris style}
			\addlegendentry{Stationary RIS}
			
			\addplot[onlymacro style]
			table[x=x, y expr = \thisrow{y}*100] from \tableDataOnlyMacro;
			
			\addplot[iab style]
			table[x=x, y expr = \thisrow{y}*100] from \tableDataIAB;
			
			\addplot[ncr style]
			table[x=x, y expr = \thisrow{y}*100] from \tableDataNCR;
			
			\addplot[ris style]
			table[x=x, y expr = \thisrow{y}*100] from \tableDataRIS;
			
			\addplot[uaviab style]
			table[x=x, y expr = \thisrow{y}*100] from \tableDataUAVIAB;
			
			\addplot[uavncr style]
			table[x=x, y expr = \thisrow{y}*100] from \tableDataUAVNCR;
		\end{axis}
	\end{tikzpicture}
	%

	\caption{\ac{CDF} of throughput in \ac{DL}.}
	\label{FIG:performance_evaluation:Throughput_down}
\end{figure}
\begin{figure}[t]
	\centering
		\begin{tikzpicture}
		\begin{axis}[common plots axis options,
			ylabel = CDF (\%),
			xlabel = Throughput (Mbps),
			ymin = 0, ymax = 100,
			xmin = -0.001, xmax = 35,
			xtick = {0,5,...,35},
			legend style={
		    	at = {(0.5, 1.05)},
		    	anchor = south,
		    	legend columns = 3,
			}
			]
			
			\def\isDownlink{False}
			
			\pgfplotstableread [col sep=comma] {\plotsDataPath/Throughput/e2e_throughput_mbps_cdf_NCR_0_backhaul_los_True_isDownlink_\isDownlink.csv}\tableDataOnlyMacro
			\pgfplotstableread [col sep=comma] {\plotsDataPath/Throughput/e2e_throughput_mbps_cdf_NCR_1_2_backhaul_los_True_isDownlink_\isDownlink.csv}\tableDataNCR
			
			\pgfplotstableread [col sep=comma] {\plotsDataPath/Throughput/e2e_throughput_mbps_cdf_IAB_1_2_backhaul_los_True_isDownlink_\isDownlink.csv}\tableDataIAB
			\pgfplotstableread [col sep=comma] {\plotsDataPath/Throughput/e2e_throughput_mbps_cdf_RIS_2_88_backhaul_los_True_isDownlink_\isDownlink.csv}\tableDataRIS
			
			\pgfplotstableread [col sep=comma] {\plotsDataPath/Throughput/e2e_throughput_mbps_cdf_UAVIAB_1_2_backhaul_los_True_isDownlink_\isDownlink.csv}\tableDataUAVIAB
			\pgfplotstableread [col sep=comma] {\plotsDataPath/Throughput/e2e_throughput_mbps_cdf_UAVNCR_1_2_backhaul_los_True_isDownlink_\isDownlink.csv}\tableDataUAVNCR

			\addlegendimage{iab style}
			\addlegendentry{Stationary IAB}
			
			\addlegendimage{ncr style}
			\addlegendentry{Stationary NCR}
			
			\addlegendimage{onlymacro style}
			\addlegendentry{Macro Only}
			
			\addlegendimage{uaviab style}
			\addlegendentry{UAV IAB}
			
			\addlegendimage{uavncr style}
			\addlegendentry{UAV NCR}
			
			\addlegendimage{ris style}
			\addlegendentry{Stationary RIS}
			
			\addplot[onlymacro style]
			table[x=x, y expr = \thisrow{y}*100] from \tableDataOnlyMacro;
			
			\addplot[iab style]
			table[x=x, y expr = \thisrow{y}*100] from \tableDataIAB;
			
			\addplot[ncr style]
			table[x=x, y expr = \thisrow{y}*100] from \tableDataNCR;
			
			\addplot[ris style]
			table[x=x, y expr = \thisrow{y}*100] from \tableDataRIS;
			
			\addplot[uaviab style]
			table[x=x, y expr = \thisrow{y}*100] from \tableDataUAVIAB;
			
			\addplot[uavncr style]
			table[x=x, y expr = \thisrow{y}*100] from \tableDataUAVNCR;
		\end{axis}
	\end{tikzpicture}
	%

	\caption{\ac{CDF} of throughput in \ac{UL}.}
	\label{FIG:performance_evaluation:Throughput_up}
\end{figure}
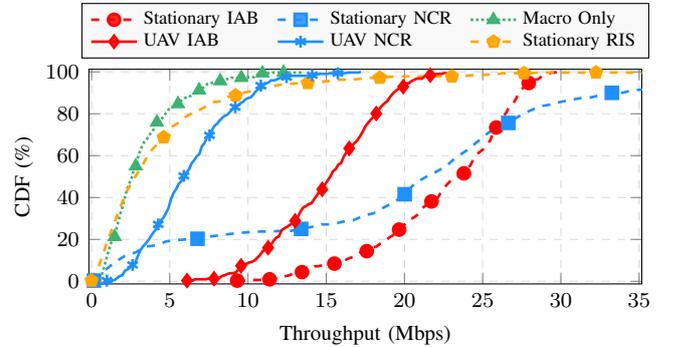

Figures~\ref{FIG:performance_evaluation:Throughput_down} and \ref{FIG:performance_evaluation:Throughput_up} present the \ac{CDF} of the \acp{UE}' throughput in \ac{DL} and \ac{UL}, respectively. %
The throughput analysis reveals similar conclusions to the \ac{SINR} results. %
The macro-only scenario, as a baseline, showed lower throughput values, especially in \ac{UL}, where 80\% of \ac{UE} connections had throughput below \SI{5}{Mbps}. %
Compared to the macro-only scenario, \ac{UE} throughput was higher with \ac{IAB}, \ac{NCR}, and \ac{RIS}. %


Comparing stationary \ac{IAB} and stationary \ac{NCR}, \acp{UE} with high throughput benefited more from \ac{NCR} due to the \ac{TDD} scheme and \ac{HD} mode of \ac{IAB}. %
In the \ac{IAB} case, a \ac{UE} required at least 2~\acp{TTI} to receive a packet, whereas with \ac{NCR}, only 1~\ac{TTI} was needed. %
Conversely, \acp{UE} with low throughput performed better with \ac{IAB} nodes. %
This is because, with \ac{IAB}, backhaul and access links are independent, whereas \ac{NCR} requires successful transmission on both links simultaneously, increasing the likelihood of retransmissions with a poor backhaul or forward link. %


In terms of throughput, \acp{UE} with low throughput saw greater improvements in \ac{DL}, while \acp{UE} with high throughput experienced higher gains in \ac{UL} from the presence of assisting nodes. %
Specifically, compared to the macro-only scenario, at the $10^{\text{th}}$ percentile, \ac{IAB} provided gains of \SI{22.12}{Mbps} (+$397.67\%$) in \ac{DL} and \SI{15.32}{Mbps} ($1,859.76\%$) in \ac{UL}. %
For the \ac{NCR}, they were \SI{17.45}{Mbps} ($317.02\%$) in \ac{DL} and \SI{0.87}{Mbps} ($114.63\%$) in \ac{UL}. %
At the $90^{\text{th}}$ percentile, they were \SI{18.05}{Mbps} ($149.50\%$) and \SI{20.86}{Mbps} ($312.42\%$), for the \ac{IAB} in \ac{DL} and \ac{IAB} in {UL}, respectively, and \SI{20.76}{Mbps} ($172.08\%$) and \SI{27.17}{Mbps} ($411.36\%$) for the \ac{NCR} in \ac{DL} and \ac{NCR} in {UL}, respectively. %
This variation is due to \ac{DL} signals being stronger at the cell edges with \ac{IAB}/\ac{NCR}, and \ac{UL} signals benefiting from shorter distances to the nodes, avoiding bottlenecks caused by low transmit power. %

Regarding \ac{RIS} throughput, the \ac{RIS} generally improved throughput for most \acp{UE}, despite not amplifying reflected signals. %
It provided an alternative path between the \ac{gNB} and \ac{UE}, helping signals bypass obstacles in the \ac{LOS} path. %
However, some \acp{UE} experienced worse throughput with \ac{RIS} compared to the macro-only scenario. %
This issue arose because, as already presented, the \ac{RIS} optimized performance for a specific cluster of \acp{UE}, and \ac{gNB} might have overestimated the \ac{SINR} for \acp{UE} less well covered by the \ac{RIS} beam, leading to failed transmissions and retries with adjusted \ac{SINR} estimates. %

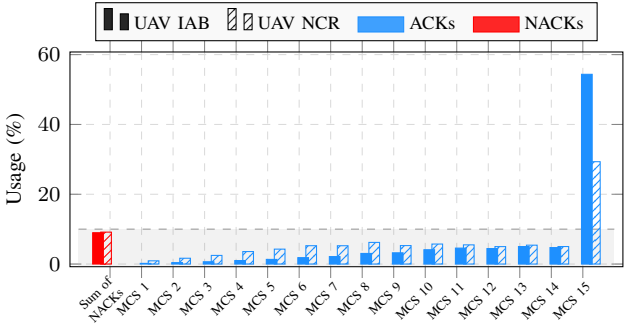
\begin{figure}[t]
	\centering
		\begin{tikzpicture}
		\begin{axis}[mcs axis options,
			ylabel=Usage (\%),
			ymax = 60,
			ybar=-1pt,
			bar width = 4pt,
			enlarge x limits={true,abs value=15pt},
			legend style={
				legend pos=north west,
				    	at = {(0.5, 1.05)},
				    	anchor = south,
				    	legend columns = 4
			}
			]
			
			\pgfplotstableread [col sep=comma] {\plotsDataPath/MCS/UAVIAB_1_2/mcs_hist_count_percentage_UAVIAB_1_2_backhaul_True_max_streams_4_LinkType_down_PicoPedestrian.csv}\tableDataIAB
			
			\pgfplotstableread [col sep=comma] {\plotsDataPath/MCS/UAVNCR_1_2/mcs_hist_count_percentage_UAVNCR_1_2_backhaul_True_max_streams_4_LinkType_down_MacroPedestrian.csv}\tableDataNCR
			
			\addlegendimage{solid bar style}
			\addlegendentry{UAV IAB}
			
			\addlegendimage{pattern bar style}
			\addlegendentry{UAV NCR}
			
			\addlegendimage{blue style}
			\addlegendentry{ACKs}
			
			\addlegendimage{red style}
			\addlegendentry{NACKs}
			
			\draw[black, opacity=0.4, fill=blue,sharp plot,dashed] (-1, 10) -- (16, 10) ;
			\fill[gray, opacity=0.1] (-1, -1) -- (-1, 10) -- (16, 10) -- (16, -1) -- cycle;
			
			\addplot[nack bar style, skip coords between index={1}{16}]
			table[x=MCS, y expr = \thisrow{nacks}*100] from \tableDataIAB;
			
			\addplot[nack bar two style, skip coords between index={1}{16}]
			table[x=MCS, y expr = \thisrow{nacks}*100] from \tableDataNCR;
			
			\addplot[ack bar style, skip coords between index={0}{1}]
			table[x=MCS, y expr =\thisrow{acks}*100] from \tableDataIAB;
			
			\addplot[ack bar two style, skip coords between index={0}{1}]
			table[x=MCS, y expr =\thisrow{acks}*100] from \tableDataNCR;

		\end{axis}
	\end{tikzpicture}
	%

	\caption{\ac{MCS} of \ac{UAV} \ac{IAB} in \ac{UL}.}
	\label{FIG:performance_evaluation:mcs_uaviab_uavncr_down}
\end{figure}
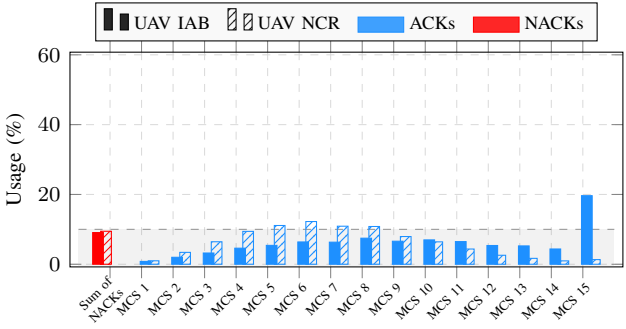
\begin{figure}[t]
	\centering
		\begin{tikzpicture}
		\begin{axis}[mcs axis options,
			ylabel=Usage (\%),
			ymax = 60,
			ybar=-1pt,
			bar width = 4pt,
			enlarge x limits={true,abs value=15pt},
			legend style={
				legend pos=north west,
				    	at = {(0.5, 1.05)},
				    	anchor = south,
				    	legend columns = 4
			}
			]
			
			\pgfplotstableread [col sep=comma] {\plotsDataPath/MCS/UAVIAB_1_2/mcs_hist_count_percentage_UAVIAB_1_2_backhaul_True_max_streams_4_LinkType_up_PicoPedestrian.csv}\tableDataIAB
			
			\pgfplotstableread [col sep=comma] {\plotsDataPath/MCS/UAVNCR_1_2/mcs_hist_count_percentage_UAVNCR_1_2_backhaul_True_max_streams_4_LinkType_up_MacroPedestrian.csv}\tableDataNCR
			
			\addlegendimage{solid bar style}
			\addlegendentry{UAV IAB}
			
			\addlegendimage{pattern bar style}
			\addlegendentry{UAV NCR}
			
			\addlegendimage{blue style}
			\addlegendentry{ACKs}
			
			\addlegendimage{red style}
			\addlegendentry{NACKs}
			
			\draw[black, opacity=0.4, fill=blue,sharp plot,dashed] (-1, 10) -- (16, 10) ;
			\fill[gray, opacity=0.1] (-1, -1) -- (-1, 10) -- (16, 10) -- (16, -1) -- cycle;
			
			\addplot[nack bar style, skip coords between index={1}{16}]
			table[x=MCS, y expr = \thisrow{nacks}*100] from \tableDataIAB;
			
			\addplot[nack bar two style, skip coords between index={1}{16}]
			table[x=MCS, y expr = \thisrow{nacks}*100] from \tableDataNCR;
			
			\addplot[ack bar style, skip coords between index={0}{1}]
			table[x=MCS, y expr =\thisrow{acks}*100] from \tableDataIAB;
			
			\addplot[ack bar two style, skip coords between index={0}{1}]
			table[x=MCS, y expr =\thisrow{acks}*100] from \tableDataNCR;

		\end{axis}
	\end{tikzpicture}
	%

	\caption{\ac{MCS} of \ac{UAV} \ac{NCR} in \ac{UL}.}
	\label{FIG:performance_evaluation:mcs_uaviab_uavncr_up}
\end{figure}

In the \ac{UAV} scenarios, \ac{UAV} \ac{IAB} outperformed \ac{UAV} \ac{NCR} in throughput. %
Compared to stationary nodes, the transmit power and gain of \ac{UAV} \ac{NCR} were insufficient to achieve high \ac{SINR}, leading \ac{UAV} \ac{NCR} to use lower \acp{MCS} than \ac{UAV} \ac{IAB} in both directions. %
This is illustrated in Figures~\ref{FIG:performance_evaluation:mcs_uaviab_uavncr_down} and \ref{FIG:performance_evaluation:mcs_uaviab_uavncr_up}, which show the \ac{MCS} and the percentage of \acp{ACK} and \acp{NACK} for \ac{UAV} \ac{IAB} and \ac{UAV} \ac{NCR} in \ac{DL} and \ac{UL}, respectively. %
\ac{ACK} and \ac{NACK} indicate successful and failed transmissions, respectively. %
The \ac{UAV} \ac{IAB} had more transmission in the \ac{MCS}~15 than the \ac{UAV} \ac{NCR} in both directions. %
Whereas, in \ac{DL}, the \ac{MCS} of \ac{UAV} \ac{NCR} was more concentrated in the \ac{MCS} 15, in \ac{UL}, its \ac{MCS} was spread to lower values. %

%

\begin{figure}[t]
	\centering
		\begin{tikzpicture}
		\begin{axis}[
			common plots axis options,
			boxplot,
			boxplot/draw direction=x,
			xlabel = Jain's fairness index,
			xmin=0.45,
			xmax=1.005,
			xtick={0.45,0.5,...,1},
			y dir=reverse,
			ytick={1,2,3,4,5,6},
			yticklabels={Only Macros,IAB,NCR,RIS,UAV IAB,UAV NCR},
		]
		
			\pgfplotstableread [col sep=comma] {\plotsDataPath/JainsIndex/raw_data_jain_index_DL.csv}\tableData
			
			\addplot[onlymacro boxplot style]
			table[y=Only Macros] from \tableData;
			
			\addplot[iab boxplot style]
			table[y=IAB] from \tableData;
			
			\addplot[ncr boxplot style]
			table[y=NCR] from \tableData;
			
			\addplot[ris boxplot style]
			table[y=RIS] from \tableData;
			
			\addplot[uaviab boxplot style] 
			table[y=UAV IAB] from \tableData;
			
			\addplot[uavncr boxplot style]
			table[y=UAV NCR] from \tableData;
			
		\end{axis}
	\end{tikzpicture}%

	\caption{Jain's fairness index in \ac{DL}.}
	\label{FIG:performance_evaluation:jain_down}
\end{figure}
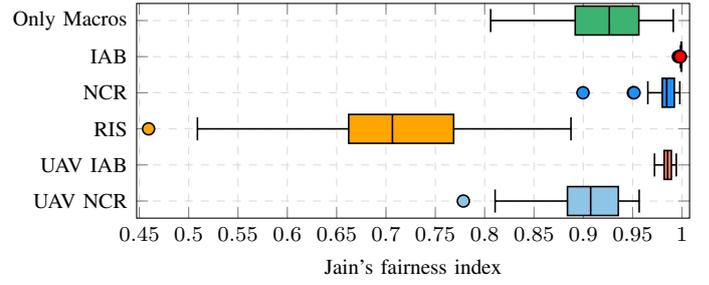

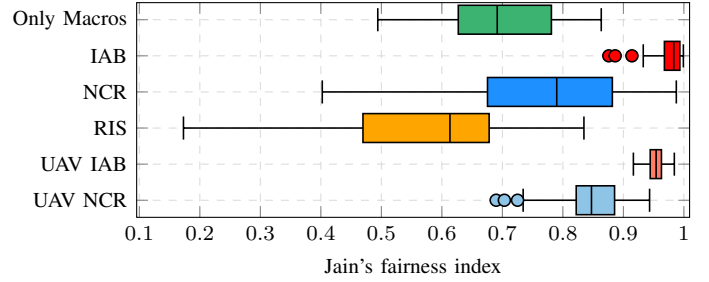
\begin{figure}[t]
	\centering
		\begin{tikzpicture}
		\begin{axis}[
			common plots axis options,
			boxplot,
			boxplot/draw direction=x,
			xlabel = Jain's fairness index,
			xmin=0.1,
			xmax=1.005,
			xtick={0.1,0.2,...,1},
			y dir=reverse,
			ytick={1,2,3,4,5,6},
			yticklabels={Only Macros,IAB,NCR,RIS,UAV IAB,UAV NCR},
			]
			
			\pgfplotstableread [col sep=comma] {\plotsDataPath/JainsIndex/raw_data_jain_index_UL.csv}\tableData
			
			\addplot[onlymacro boxplot style]
			table[y=Only Macros] from \tableData;
			
			\addplot[iab boxplot style]
			table[y=IAB] from \tableData;
			
			\addplot[ncr boxplot style]
			table[y=NCR] from \tableData;
			
			\addplot[ris boxplot style]
			table[y=RIS] from \tableData;
			
			\addplot[uaviab boxplot style] 
			table[y=UAV IAB] from \tableData;
			
			\addplot[uavncr boxplot style]
			table[y=UAV NCR] from \tableData;
			
		\end{axis}
	\end{tikzpicture}%

	\caption{Jain's fairness index in \ac{UL}.}
	\label{FIG:performance_evaluation:jain_up}
\end{figure}

Finally, Figures~\ref{FIG:performance_evaluation:jain_down} and~\ref{FIG:performance_evaluation:jain_up} present the Jain's fairness index in terms of \acp{UE}' throughput in \ac{DL} and \ac{UL}, respectively. The Jain's fairness index, i.e., $\Delta = (\sum_{u \in \stU} \theta_{u})^2/(|\stU| \sum_{u \in \stU} (\theta_{u})^2)$ with $\theta_u$ being the throughput of UE $u$, is defined as the ratio between the square of the sum of all \acp{UE}' throughputs and the product between the number of \acp{UE} and the sum of the squares of each \acp{UE}' throughput~\cite{Jain1998}. %
\ac{RIS} had the lowest fairness between the \acp{UE}. %
This is due to the fact that \ac{RIS} implementation benefited just a set of \acp{UE} to which the signal reflected by \ac{RIS} highly improved the \acp{UE}' \ac{SINR}, while, the other \acp{UE} in the system continued with low \ac{SINR} being served by the \ac{gNB}. %
In the \ac{IAB} case, as explained before, \acp{UE} with high \ac{SINR} gain had their gain in terms of throughput limited by the \ac{TDD} schemes and \ac{HD} mode (at least 2 \acp{TTI} to execute a transmission). %
This made the system fairer, by avoiding some \acp{UE} to have a really high throughput gain. %

\section{Conclusions}
\label{SEC:CONC}

In light of the growing need for network densification, \ac{IAB}, \ac{NCR}, and \ac{RIS} are alternative technologies to assist the \acp{gNB}. %
These nodes can be more cost-effective and quicker to deploy than wired backhaul solutions. %
In this context, the present work overviewed their main characteristics also providing their main investigated topics. %
Moreover, their performances were analyzed in the context of an outdoor sport event. %
The study compared stationary deployments of \ac{IAB}, \ac{NCR}, and \ac{RIS}, as well as mobile \ac{IAB} and \ac{NCR} mounted on \acp{UAV}. %
The analyses concluded that both stationary and mobile (UAV-mounted) \ac{IAB} and \ac{NCR} deployments are effective for extending coverage to cell-edge and obstructed areas. %
Also, both \ac{IAB} and \ac{NCR} outperform the \ac{RIS}, in terms of throughput and \ac{UE}'s experienced \ac{SINR}.
They also improve the \acp{UE} \ac{UL} transmissions, which can be a bottleneck in traditional networks. %
Moreover, these solutions also facilitate the rapid deployment of new cells, enhancing connectivity in high-demanding areas. %
Furthermore, \ac{UAV}-mounted nodes have proven to be useful in scenarios where \ac{LOS} links for stationary deployments are obstructed. %
However, successful implementation of these technologies requires careful network planning and environment-specific topology design. %

\printbibliography{}
\endrefsection{}
\ifCLASSOPTIONcaptionsoff
	\newpage
\fi

\end{document}